\documentclass{article}
\usepackage[a4paper, portrait, margin=1.1811in]{geometry}
\usepackage[english]{babel}
\usepackage[utf8]{inputenc}
\usepackage[T1]{fontenc}
\usepackage{helvet}
\usepackage{etoolbox}
\usepackage{graphicx}
\usepackage{url}
\usepackage{titlesec}
\usepackage{amssymb}
\usepackage{amsthm}
\usepackage{mathtools}
\usepackage{subcaption} 
\usepackage{tabularx}
\usepackage{colortbl}
\usepackage{xcolor}
\usepackage{hhline}
\usepackage{float}
\usepackage{times}
\usepackage{listings}
\usepackage{setspace}
\usepackage{enumerate}
\usepackage{ulem}
\usepackage{textcomp}
\usepackage{stfloats}
\usepackage{verbatim}
\usepackage{booktabs}
\usepackage{mathrsfs}
\usepackage{multirow}
\usepackage{multicol}
\usepackage{subcaption}
\usepackage{adjustbox}
\usepackage{algorithm}
\usepackage{algpseudocode}
\usepackage{tikz}
 \usetikzlibrary{calc}
\usetikzlibrary{arrows.meta,
                chains,
                positioning,
                shapes.geometric,matrix, backgrounds}
\makeatletter
\patchcmd{\@maketitle}{\LARGE \@title}{\fontsize{16}{19.2}\selectfont\@title}{}{}
\makeatother

\usepackage{authblk}

\newsavebox\affbox
\author{Siyi~Wang, Alexandre~Leblanc, Paul~D.~McNicholas}

\titleformat{\section}{\normalfont\fontsize{10}{15}\bfseries}{\thesection.}{1em}{}
\titleformat{\subsection}{\normalfont\fontsize{10}{15}\bfseries}{\thesubsection.}{1em}{}
\titleformat{\subsubsection}{\normalfont\fontsize{10}{15}\bfseries}{\thesubsubsection.}{1em}{}

\title{Automatic depth-based local center clustering via $\beta$-integrated local depth and adaptive grouping}

\date{}    

\begin{document}

\pagestyle{headings}	
\newpage
\setcounter{page}{1}
\renewcommand{\thepage}{\arabic{page}}
\newtheorem{thm}{\bf Definition}[section]
\newtheorem{remark}{\bf Remark}[section]
\newtheorem{pro}{\bf Proposition}[section]
\newtheorem{theorem}{\bf Theorem}[section]
\newcommand{\diag}{\text{diag}}
\newcommand{\tr}{\text{tr}}
\newcommand{\load}{\mathbf\Lambda}
\newcommand{\noisev}{\mathbf\Psi}
\newcommand{\Beta}{\mbox{\boldmath$\beta$}}
\newcommand{\sampcov}{\mathbf{S}}
\newcommand{\ident}{\mathbf{I}}
\newcommand{\vecx}{\mathbf{x}}
\newcommand{\vece}{\mathbf{e}}
\newcommand{\vecE}{\mathbf{E}}
\newcommand{\vecX}{\mathbf{X}}
\newcommand{\vecU}{\mathbf{U}}
\newcommand{\vecR}{\mathbf{R}}
\newcommand{\vecQ}{\mathbf{Q}}
\newcommand{\vecV}{\mathbf{V}}
\newcommand{\vecS}{\mathbf{S}}
\newcommand{\vecN}{\mathbf{N}}
\newcommand{\vecG}{\mathbf{G}}
\newcommand{\vecC}{\mathbf{C}}
\newcommand{\vecw}{\mathbf{w}}
\newcommand{\vecz}{\mathbf{z}}
\newcommand{\vecy}{\mathbf{y}}
\newcommand{\vecu}{\mathbf{u}}
\newcommand{\vecc}{\mathbf{c}}
\newcommand{\veca}{\mathbf{a}}
\newcommand{\vecg}{\mathbf{g}}
\newcommand{\vecv}{\mathbf{v}}
\newcommand{\bbS}{\mathbb{S}}
\newcommand{\mcD}{\mathscr{D}}
\newcommand{\tbbG}{\Tilde{\mathbb{G}}}
\newcommand{\tbbS}{\Tilde{\mathbb{S}}}
\newcommand{\vecnu}{\mathbf\nu}
\newcommand{\thet}{\mathbf{\Theta}}
\newcommand{\new}{\mbox{\tiny new}}
\newcommand{\vecmu}{\boldsymbol\mu}
\newcommand{\veclam}{\mathbf\lambda}
\newcommand{\vecLambda}{\mathbf\Lambda}
\newcommand{\varthet}{\boldsymbol\vartheta}
\newcommand{\vecSigma}{\mathbf\Sigma}
\newcommand{\vecXi}{\mathbf\Xi}
\newcommand{\vecPsi}{\mathbf\Psi}
\newcommand{\vecBeta}{\mathbf\Beta}
\newcommand{\vecTheta}{\mathbf\Theta}
\newcommand{\vecDelta}{\mathbf\Delta}
\newcommand{\vectheta}{\boldsymbol\theta}
\newcommand{\vecepsilon}{\boldsymbol\epsilon}
\newcommand{\vecdelta}{\boldsymbol\delta}
\newcommand{\vecpi}{\boldsymbol\pi}
\newcommand{\vecalpha}{\boldsymbol\alpha}
\newcommand{\vecvartheta}{\boldsymbol\vartheta}
\newcommand{\argmin}{\arg\min}
\newcommand{\argmax}{\arg\max}
\newtheorem{assumption}{Assumption}

\captionsetup[figure]{labelfont={bf},labelformat={default},labelsep=period,name={Figure }}	\captionsetup[table]{labelfont={bf},labelformat={default},labelsep=period,name={Table }}
\setlength{\parskip}{0.5em}
	
\maketitle
\noindent\rule{15cm}{0.5pt}
	\begin{abstract}
    Clustering is an unsupervised learning technique that partitions unlabeled data into groups. Most existing methods require user-specified parameters, such as the number of clusters or neighborhood size. Conversely, we propose automatic depth-based local center clustering (A-DLCC), a fully data-driven method that eliminates numerical parameter tuning. A-DLCC uses the $\beta$-integrated local depth to identify stable exemplars, points consistently central across multiple locality levels, termed local centers, which are ranked by their representativeness. Each local center induces a group of similar points, with group-level similarity measured by a proposed nonparametric metric called group-level local similarity. To guide merging, we incorporate the bottleneck path idea from graph theory, which forms the basis of our adaptive merging criterion. Based on this criterion, we design a single agglomeration rule in which a group is either absorbed by a neighbor it reaches better than itself or bonded to a neighbor that both sides find more reachable than their own background, every merge being additionally required to be carried by a contact stronger than a configuration-model null expects. The rule automatically estimates the number of clusters and decides when to stop merging. Experiments on synthetic and real data show that A-DLCC produces interpretable clustering results without parameter tuning.
		 \\ \\
		\let\thefootnote\relax\footnotetext{
			
		}
	\end{abstract}
\noindent\rule{15cm}{0.4pt}

\section{Introduction}
\algrenewcommand\algorithmicrequire{\textbf{Input:}}
\algrenewcommand\algorithmicensure{\textbf{Output:}}
Clustering is a central task in unsupervised learning, with applications in representation learning \cite{liu2023dinosr}, pattern recognition \cite{diday1981clustering}, image analysis \cite{mittal2022comprehensive} and other natural sciences \cite{kisi2025integration}. The primary objective of clustering is to produce meaningful and interpretable partitions of data without annotated class labels. Clustering is therefore a standard tool for data analysis, especially when little or no prior knowledge about the data is available.

Many clustering algorithms have been proposed, ranging from classic methods such as $K$means~\cite{lloyd1982least}, hierarchical clustering~\cite{johnson1967hierarchical}, density-based approaches \cite{ester1996density, rodriguez2014clustering} and Gaussian mixture models (GMM)~\cite{fraley2002model}, to more recent approaches based on graph theory \cite{ng2001spectral} and manifold learning \cite{souvenir2005manifold}. As real-world data become larger and more heterogeneous~\cite{jain2010data}, many algorithms require tuning multiple parameters or careful model selection to achieve satisfactory performance.

In practice, model and parameter selection are often performed by running the clustering algorithm repeatedly with different settings, then choosing the best result according to some criterion, such as silhouette width~\cite{rousseeuw1987silhouettes} for $K$means or the Bayesian information criterion (BIC) \cite{schwarz78} for model-based clustering. However, methods that can estimate the number of clusters automatically and deliver high-quality results without the need for parameter selection remain relatively limited. 
\subsection{Literature review}
Many approaches have been proposed for parameter-free or adaptive clustering. Density-based clustering algorithms are often considered adaptive, as they do not require a pre-specified number of clusters and can detect clusters of arbitrary shape and size. However, they typically depend on other parameters, such as neighborhood size or density thresholds. Mean-shift~\cite{fukunaga1975estimation} is a classic example, which iteratively shifts data points toward regions of higher density, effectively discovering clusters without requiring $K$ as input. However, mean-shift relies on a kernel bandwidth parameter, which critically affects its performance. Recent developments have focused on improving efficiency and sensitivity to parameter choices, for example by using $k$-nearest neighbor ($k$NN) density estimates instead of fixed bandwidths, as in NN-robust mean-shift (NN-RMS) \cite{cariou2022novel} and NN-blurring mean-shift \cite{beck2019distributed}. Nevertheless, mean-shift-based methods still require users to specify parameters that control density estimation.

FINCH~\cite{sarfraz2019efficient} is a representative example of hierarchical clustering that aims to produce clustering results automatically. At each iteration, it merges all mutually closest groups (or those connected by a path), generating a hierarchy of clusterings at multiple levels of granularity without requiring any parameter input. However, the final selection from the resulting hierarchy still depends on user decisions or post hoc evaluation.

For methods that rely on interpretable parameters, such as the number of clusters $K$, rule-of-thumb heuristics are commonly employed. The eigengap heuristic in spectral clustering~\cite{von2007tutorial} is a well-known example, but its reliability diminishes for non-Gaussian cluster structures, as demonstrated by John et al.~\cite{john2020spectrum}. To address this, they proposed the spectrum method, introducing the multimodality gap for selecting $K$ in non-Gaussian clusters, and further enhancing the process by employing a density-aware kernel and GMM on the eigenvector matrix instead of $K$means. Another direction in automatic spectral clustering focuses on adaptively selecting the affinity matrix, which can also be regarded as a key parameter. For example, Fan et al.~\cite{fan2022simple} proposed using the relative eigengap or Bayesian optimization to automatically select both the affinity matrix and its hyperparameters from candidate options. However, their approach still requires the number of clusters to be specified in advance, and thus cannot simultaneously determine both the optimal $K$ and the affinity matrix.

A theoretically principled approach to automatic clustering is to define an objective function and seek the partition that optimizes it. Most methods in this category are extensions of $K$means, accepting split or merge operations that most improve the objective in each iteration. Early examples include X-means \cite{dan2000extending}, which optimizes the BIC but still requires specifying a range of cluster numbers in advance. More recent developments address this limitation. For instance, unsupervised $K$means (U-$K$means) \cite{sinaga2020unsupervised} maximizes an entropy-based objective via an expectation-maximization-like algorithm, with the estimated number of clusters stabilizing over iterations. $K^*$means \cite{mahon2025k} formulates the objective using the minimum description length principle. This approach, especially when combined with UMAP~\cite{mcinnes2018umap} for dimensional reduction, has performed well in estimating $K$ even for large datasets. However, these criteria are generally most effective for clusters with convex structure, and there is still a lack of widely applicable and computationally efficient objective function that generalizes to more complex cluster shapes. 
\subsection{Motivation and contribution}
Despite various efforts toward automatic clustering, most existing methods still require parameter tuning, user intervention, or rely on strong assumptions about cluster structure—often reflecting the limits of the clustering algorithms themselves.  Depth-based local center clustering (DLCC)~\cite{wang2025depth} is a recently proposed method that uses statistical depth to address a broad range of clustering challenges, including non-convex shapes, unbalanced sizes, overlapping clusters, and high-dimensional data. However, DLCC itself requires the selection of a neighborhood size and similarity threshold to determine and group local centers (i.e., points that are locally central within a subset of the data), which limits its usability in practice without prior knowledge. This motivates the development of a fully automatic version of DLCC that removes assumptions on cluster size and number. The new method keeps the flexibility of DLCC and removes the need for tuning parameters.

Several obstacles arise when attempting to remove parameters from DLCC. The neighborhood size parameter determines definitions of both neighborhoods and local centers, and thus influences the similarity between neighborhoods as well. The threshold parameter governs how local centers are grouped. Eliminating parameter tuning thus requires to solve the following main challenges: first, how to identify exemplars (local centers) without relying on a fixed parameter for neighborhood size; second, how to define a nonparametric measure of similarity between these exemplars, focusing on the similarity between their associated point groups rather than between individual points; and third, how to design an algorithm that can automatically determine the number of clusters or provide an appropriate stopping rule for merging. To address these challenges, we propose automatic depth-based local center clustering (A-DLCC). Our approach consists of the following main contributions.
\begin{itemize}
    \item Building on $\beta$-integrated local depth ($\beta$-ILD; \cite{wang2025beta}), we introduce a parameter-free method for defining exemplars—points that consistently occupy central positions within local neighborhoods of varying size. We also establish a representativeness ranking, where more representative points are those most often identified as locally central, particularly in larger neighborhoods. This ranking may have broader applications, such as centroid initialization in other clustering methods.
    \item Inspired by partitioned local depth (PaLD; \cite{berenhaut2022social}), we propose a nonparametric measure for quantifying similarity between disjoint groups of points, termed group-level local similarity (GLS). This measure inherits desirable properties from PaLD, such as naturally assigning zero similarity to completely separated groups and invariance under similarity transformations (i.e., rotation, dilation, and shift).
    \item We define intra-group reachability, inter-group reachability, and the relative reachability ratio, and design an adaptive merging criterion based on these measures to guide group aggregation. The criterion is applied through one agglomeration rule with two directional modes of merging (absorption and bonding), a null-model test of the contact between the groups on the depth graph, and a final reconsideration of small groups. The same rule covers both the connected-shape and the well-separated scenarios.
\end{itemize}
These contributions may also be useful for exemplar selection, group similarity measurement, and the design of merging criteria in other settings.

The remainder of the paper is organized as follows. Section~\ref{sec:pre} reviews essential concepts, including spatial depth, $\beta$-ILD, and DLCC. Section~\ref{sec:pointrep} presents our approach for defining local centers and point representativeness using $\beta$-ILD. Section~\ref{sec:adlcc} details the A-DLCC algorithm, including group-level similarity, adaptive merging, and other modifications from the original DLCC for a fully automatic framework. Section~\ref{sec:app} demonstrates that A-DLCC achieves satisfactory and interpretable results on a wide variety of both synthetic and real datasets. Finally, Section~\ref{sec:cfd} concludes the paper and outlines potential future directions.
\section{Preliminaries} \label{sec:pre}
We review the concepts from data depth used in A-DLCC. 
\subsection{Local depth and integrated local depth}
A depth function provides a center-outward ordering of data points and generalizes univariate order statistics such as the median, quantiles, and ranks to higher dimensions \cite{mozharovskyi2022data}. In clustering, early studies mainly used data depth as an auxiliary tool—computing depth within each cluster to refine partitions. More recently, local depth concepts have been used to define exemplars directly. For instance, Francisci et al.~\cite{francisci2023analytical} adopt $\tau$-local depth~\cite{agostinelli2011local} for mean-shift-like clustering, and DLCC~\cite{wang2025depth} uses $\beta$-local depth ($\beta$-LD)~\cite{paindaveine2013depth} to construct the similarity matrix and define local centers.

\begin{figure}[h]
\centering
\includegraphics[width=0.45\linewidth]{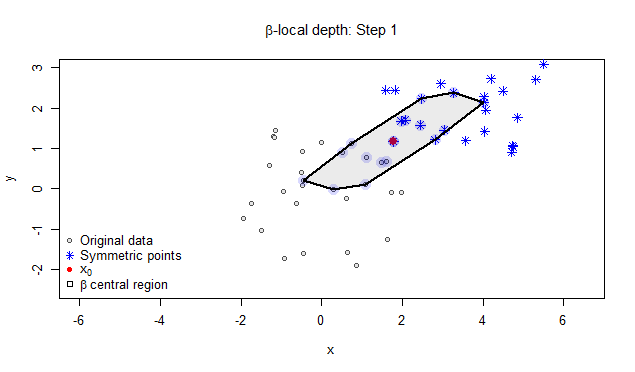}\hfill
\includegraphics[width=0.45\linewidth]{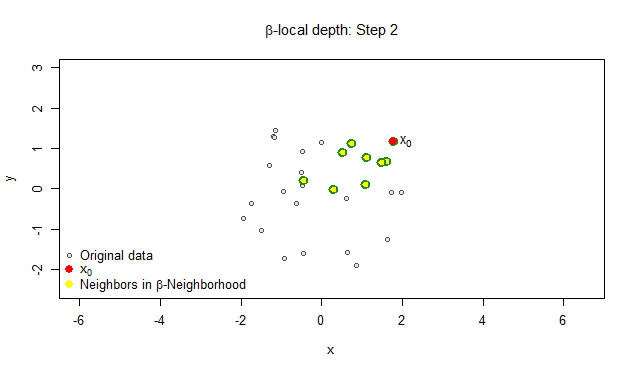}
\caption{$\beta$-local halfspace depth construction for a sample of size $n=30$, with $\beta=0.3$. (a) Construction of the $\beta$-local depth region for $\vecx_0$. (b) Depth-defined neighbors for $\vecx_0$. The local depth of $\vecx_0$ is computed w.r.t. these neighbors.}
\label{fig:beta_local}
\end{figure}

Figure~\ref{fig:beta_local} illustrates the construction of sample $\beta$-LD. Given a point $\vecx_0$ and a dataset $\mathbf{X}$ with $n$ observations, we first create the reflected set $\mathbf{X}_{R\vecx_0} = \{\mathbf{X} \cup {2\vecx_0 - \vecx_j : \vecx_j \in \mathbf{X}}\}$, ensuring central symmetry at $\vecx_0$. The $\beta$-neighborhood of $\vecx_0$ is defined as the $\lceil n\beta \rceil$ points in $\mathbf{X}$ with the highest depth values with respect to (w.r.t.) $\mathbf{X}_{R\vecx_0}$ (highlighted in yellow in Figure~\ref{fig:beta_local}). The sample $\beta$-LD of $\vecx_0$ is defined as the depth of $\vecx_0$ relative to its $\beta$-neighborhood:
\begin{eqnarray}
    \mathrm{LD}^\beta(\vecx_0\mid \vecX)=D(\vecx_0\mid \vecN_{\vecx_0}^\beta),
\end{eqnarray}
where $D$ denotes the chosen depth function and $\vecN_{\vecx_0}^\beta$ is the $\beta$-neighborhood of $\vecx_0$.

$\beta$-ILD extends the idea of $\beta$-LD by integrating local depth values over a range of locality levels $\beta \in (0,1]$. This integration leads to the resulting depth capturing both local and global structures in the data without fixing a particular $\beta$. 
\begin{thm}[Sample $\beta$-integrated local depth]\label{def:BILD}
Given a dataset $\vecX = \{\vecx_1, \ldots, \vecx_n\}$, the sample $\beta$-ILD for a point $\vecx$ is defined as
\begin{eqnarray}
    \mathrm{ILD}^{\beta_b}(\vecx\mid\vecX)=\sum_{i=1}^{b}\mathrm{LD}^{\beta_{i}}(\vecx\mid \vecX) \int_{\beta_{i-1}}^{\beta_{i}} w(\beta) \, d\beta,
\end{eqnarray}
where $\beta_{i+1} = \beta_i + 1/n$, and $w(\beta)$ is a weighting function satisfying $\int_{\beta_0}^{\beta_b} w(\beta)d\beta = 1$.
\end{thm}
In practice, $\beta_0$ is set to a small positive value so that local depth is well-defined for any sample size larger than $n\beta_0$ (this paper sets $\beta_0 = \min(\frac{2d-1}{n}, \frac{9}{n})$, where $d$ is the dimension of the data, to avoid instability of local depth values in very small neighborhoods), and $\beta_b = 1$. The weighting function is taken to be the uniform distribution between $\beta_0$ and $\beta_b$.

Although many notions of data depth have been proposed, most are computationally intensive in high dimensions (e.g., halfspace depth~\cite{Tukey1975} and projection depth~\cite{zuo2000general} require $O(n^d)$, simplicial depth~\cite{liu1990notion} requires $O(n^{d+1})$; see~\cite{mosler2022choosing} for a comprehensive overview). For this reason, DLCC adopts spatial depth~\cite{serfling2002depth}, which can be computed in $O(nd)$ time per point. Since A-DLCC relies on $\beta$-ILD, which requires repeated local depth computations across multiple locality levels, it also uses spatial depth for computational efficiency. The spatial depth for a point $\vecz$ with respect to a sample $\mathbf{X}$ is
\begin{equation}
D_{\mathrm{SD}}(\vecz\mid \mathbf{X})=
1-\left\| \sum_{i=1}^n \frac{\vecz-\vecx_i}{n\| \vecz-\vecx_i \|} \right\|.
\label{eq:sd}
\end{equation}
\subsection{DLCC algorithm}
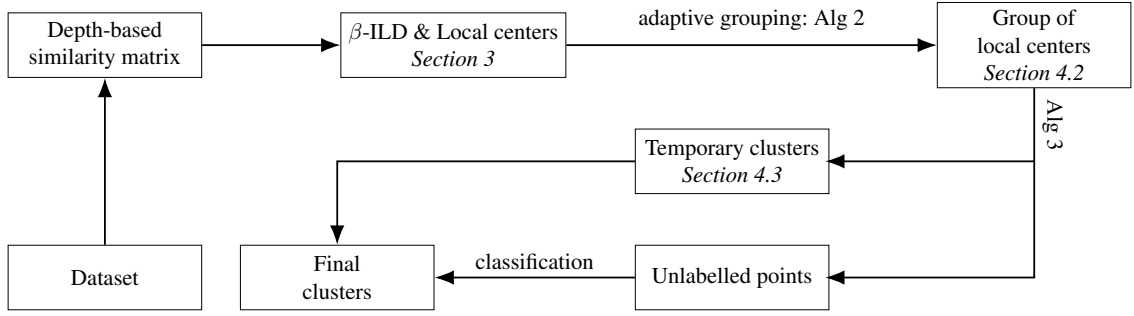
\begin{figure*}[!ht]
\centering
\resizebox{\textwidth}{!}{
\begin{tikzpicture}[scale=1.8,
  box/.style={draw, rectangle, minimum width=3cm, minimum height=1cm, align=center},
  arr/.style={-{Latex[length=3mm]}, thick}
]

  \node (dataset) at (0,-2)   [box] {Dataset};
  \node (depth)   at (0, 0)   [box] {Depth-based\\similarity matrix};
  \node (beta)    at (3, 0)   [box] {$\beta$-ILD \& Local centers\\\textit{Section~\ref{sec:pointrep}}};
  \node (group)   at (8, 0)   [box] {Group of\\local centers\\\textit{Section~\ref{sec:ags}}};
  \coordinate (drop)   at (8,-1);
  \coordinate (corner) at (8,-2);
  \coordinate (tmpc) at (6.2,-1);
  \coordinate (ulabel) at (6.2,-2);
  \node (temp)    at (5.4,-1) [box] {Temporary clusters\\\textit{Section~\ref{sec:tc}}}; 
  \node (unlab)   at (5.4,-2)   [box] {Unlabelled points}; 
  \node (final)   at (2,-2)   [box] {Final\\clusters};

  \draw[arr] (dataset.north) -- (depth.south);

  \draw[arr] (depth.east) -- (beta.west);

  \draw[arr] (beta.east) -- node[above,yshift=0.1cm]{adaptive grouping: Alg~\ref{alg:reach}}
                       (group.west);
  \draw[arr]   (group.south) -- node[right,pos=0.5]{\rotatebox{270}{Alg~\ref{alg:utc}}} 
  (drop) -- (tmpc);
  \draw[arr] (drop) --(corner) -- (ulabel);
   \coordinate (tmpcor) at (2,-1);
  \draw[arr] (temp.west) --(tmpcor)-- (final.north);
  \draw[arr] (unlab.west) -- node[above,midway]{classification} (final.east);
\end{tikzpicture}
}
\caption{Flowchart for the A-DLCC algorithm}
\label{fig:flowchartADLCC}
\end{figure*}

With the basic concepts of depth and local depth in place, we briefly review the DLCC framework, which consists of the following steps:
\begin{enumerate} [Step 1]
\item Construct the depth-based similarity matrix $\bbS$, where each entry is defined as $\bbS_{i,j} = D(\vecx_j \mid \mathbf{X}_{R\vecx_i})$. In both DLCC and A-DLCC, symmetry is enforced by averaging $\bbS_{i,j}$ and $\bbS_{j,i}$. Given a neighborhood size parameter (or equivalently, $\beta$), define each point’s neighborhood accordingly.
\item For each neighborhood identify the deepest point, and filter these candidates by additional criteria. The filtered local centers are then grouped using either the ``min'' or ``max'' strategies, both of which allow a cluster to have multiple exemplars.
\item The resulting local center groups determine the number of clusters. Remaining points are assigned to clusters based on their similarity scores to each group and whether they are included in the neighborhoods of the group’s local centers. This produces the temporary clusters and ambiguous points that remain unlabeled, because of a lack of similarity to any local center.
\item Finally, classification techniques such as $k$NN, random forest, or the maximal depth classifier are applied to assign the unlabeled points to an existing cluster.
\end{enumerate}
Figure~\ref{fig:flowchartADLCC} shows the flowchart of the A‑DLCC algorithm. While the overall workflow follows that of the original DLCC, certain steps indicated by corresponding Section or Algorithm labels (i.e., defining and grouping local centers and constructing temporary clusters) have been modified to eliminate the need for predefined parameters. The Python implementation is available at GitHub\footnote{\url{https://github.com/lytgysrn/ADLCC-python}}.

\subsection{Summary of notations}
To improve clarity, we summarize in Table \ref{tab:notation} the main notations used in subsequent sections.
\begin{table}[ht]
\centering
\caption{Summary of notation.}\label{tab:notation}
\small
\begin{minipage}[t]{0.48\textwidth}
\centering
\begin{tabularx}{\linewidth}{@{}lX@{}}
\toprule
Notation & Description \\
\midrule
$\vecX$           & Dataset, $\{\vecx_1, \ldots, \vecx_n\}$ \\
$n$, $d$          & Number of observations and dimensions of $\vecX$ \\
$\beta$, $b$       & Locality level and number of locality levels \\
$\mathrm{LD}^\beta(\vecx)$, $\mathrm{ILD}^\beta(\vecx)$  
                  & $\beta$-LD and $\beta$-ILD at $\vecx$ \\
$\vecN_\vecx^\beta$& $\beta$-neighborhood of point $\vecx$ \\
$f$                & Frequency \\
$\vecc$, $\vecC$            & local center and set of local centers\\
$\mathcal{E}$, $\mathcal{E}'$ 
                   & Sets of locally deep points; $\mathcal{E}' = \mathcal{E} \setminus \vecC$ \\
$\bbS$, $\mathrm{DS}(\cdot,\cdot)$      & Depth-based similarity matrix, and similarity between two points, i.e., $\bbS_{i,j} = \mathrm{DS}(\vecx_i, \vecx_j)$ \\
$\vecG_t$          & Group of points assigned to local center $\vecc_t$ \\
$\mathrm{GLS}(i,j)$& Initial group-level similarity between $\vecG_i$ and $\vecG_j$ \\
$\mathrm{rs}_{\vecG}(\cdot,\cdot)$ & Reachable similarity within group $\vecG$ \\
\bottomrule
\end{tabularx}
\end{minipage}
\hfill
\begin{minipage}[t]{0.48\textwidth}
\centering
\begin{tabularx}{\linewidth}{@{}lX@{}}
\toprule
Notation & Description \\
\midrule
$\mathbb{G}$       & Group similarity matrix \\
$\mu_{\vecG}(\vecx)$ & Intra-group similarity of $\vecx$ in $\vecG$ \\
$\mu_{\vecG_i \to \vecG_j}(\vecx)$ & Between-group similarity from $\vecx \in \vecG_i$ to group $\vecG_j$ \\
$\rho_{\vecG_i \to \vecG_j}$ & Relative reachability ratio from $\vecG_i$ to $\vecG_j$ \\
$\tbbS$, $\tbbG$ & Reachable versions of $\bbS$ and $\mathbb{G}$ \\
$\bar\phi_i$, $\phi_i$ & Background level of $\vecG_i$ and its clipped version (base threshold) \\
$\omega_{i|j}$ & Disruption suffered by $\vecG_i$ when merged with $\vecG_j$ \\
$Q_{\mathrm{th}}(i\mid j)$  & Adaptive threshold applied by $\vecG_i$ to the merge with $\vecG_j$ \\
$\Delta Q_{ij}$ & Modularity gain of joining $\vecG_i$ and $\vecG_j$ on the depth graph \\
$\vecg_k$          & A group of local centers \\
$\vecG_{\vecg}$, $\mathrm{b}(\vecg)$ & Points of a group of local centers, and the weakest similarity its connectivity relies on \\
$\mathcal{T}_k$    & Temporary cluster corresponding to $\vecg_k$ \\
\bottomrule
\end{tabularx}
\end{minipage}

\end{table}

\section{$\beta$-integrated local depth-based point representativeness} \label{sec:pointrep}
We propose a nonparametric method to identify exemplars within a dataset, together with a representativeness ranking. Recall the $\beta$-ILD definition in Section~\ref{sec:pre}. For a dataset $\vecX = \{ \vecx_1, \dots, \vecx_n \}$, define a sequence of locality levels 
$\beta_1 < \cdots < \beta_b$ with $\beta_b = 1$ and $\beta_{i+1} = \beta_i + 1/n$. 
At each level $\beta_j$, we compute the local depth $\mathrm{LD}^{\beta_j}(\vecx_i)$ 
for every data point $\vecx_i$. Collect these values into an $n \times b$ matrix $\mathbb{L}$, with entry $(i,j)$ given by $\mathrm{LD}^{\beta_j}(\vecx_i)$. 

Similarly, for $\beta$-ILD with a uniform weighting function across locality levels, a matrix storing ILD values can be constructed, denoted $\mathbb{IL}$. Specifically, for each row $i$ in $\mathbb{L}$, the ILD values are computed as
$$
\mathbb{IL}_{i,j} = \frac{1}{j} \sum_{u=1}^{j} \mathbb{L}_{i,u}, \quad j = 1, \dots, b. 
$$
\begin{thm}[Locally deep point]
For each data point $\vecx_i$ and locality level $\beta_j$, consider the neighborhood $\vecN_{\vecx_i}^{\beta_j}$. The locally deep point for the subset defined by $\vecN_{\vecx_i}^{\beta_j}$ is the point $\vecz \in \vecN_{\vecx_i}^{\beta_j}$ that maximizes $\mathrm{ILD}^{\beta_j}(\vecz)$.
\end{thm}

It is important to note that for any point \( \vecz \in \vecN_{\vecx_i}^{\beta_j} \), the ILD value is computed based on the neighborhood \( \vecN_{\vecz}^{\beta_j} \), rather than the reference neighborhood \( \vecN_{\vecx_i}^{\beta_j} \). Specifically, when determining the locally deep point of \( \vecN_{\vecx_i}^\beta \), we compare the values in the $j$-th column of the matrix \( \mathbb{IL} \). 

Applying the above definition to each of the $nb$ subsets (one for each $\vecx_i$ and each $\beta_j$) yields a collection of candidate local centers. Naturally, the frequency of each candidate local center can be defined.
\begin{thm}[Frequency of locally deep point]
The \emph{frequency} $f$ of a locally deep point $\vecz$ is defined as the number of subsets across all locality levels and points in which $\vecz$ emerges as the ILD-based local center.
\end{thm}
\begin{remark}\label{rem:E}
We denote by $\mathcal{E}$ the ordered set of all locally deep points sorted by decreasing frequency, and by $\mathcal{E}' \subseteq \mathcal{E}$ those not selected as local centers in Section~\ref{sec:slc}. The set $\mathcal{E}'$ is referred to in Section~\ref{sec:tc}.
\end{remark}
\begin{figure}[t]
\centering
\includegraphics[width=0.45\linewidth]{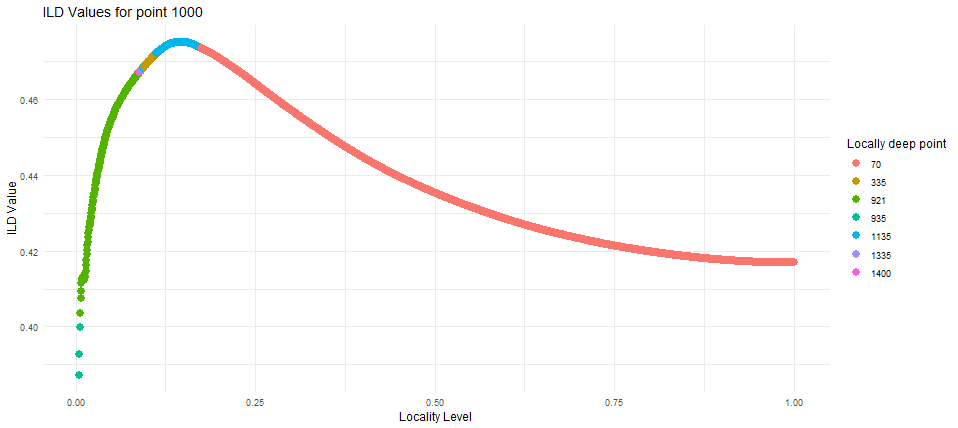}\hfill
\includegraphics[width=0.45\linewidth]{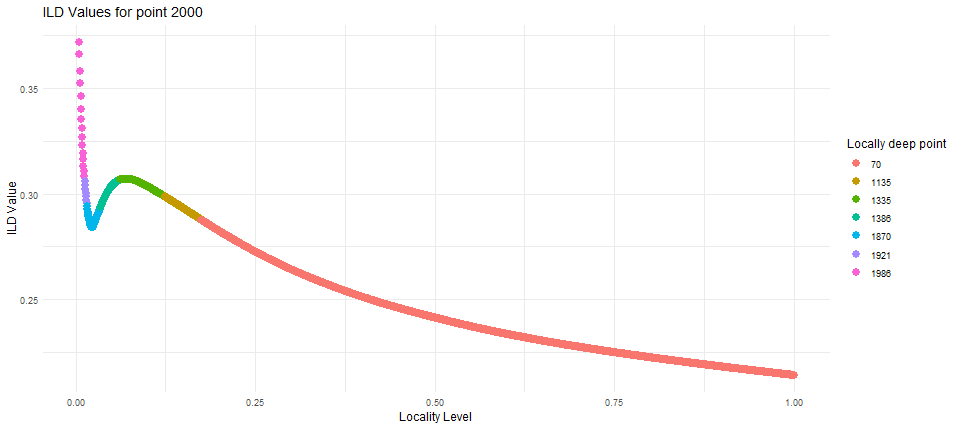}
\caption[ILD values for two chosen points (Point $1000$ and Point $2000$) in the Yale-B dataset (Section \ref{sec:rd}) across varying locality levels.]{ILD values for two chosen points (Point $1000$ and Point $2000$) in the Yale-B dataset (Section \ref{sec:rd}) across varying locality levels. Colors indicate the locally deepest point in each neighborhood.}
\label{fig:ILDexample}
\end{figure}
Compared to $\beta$-LD, the $\beta$-ILD provides a smoothed summary of local depth values, reducing fluctuations and thereby avoiding the identification of an excessive number of locally deep points. As higher locality levels correspond to larger neighborhoods that naturally include more points, any given point is more likely to appear in the neighborhoods of others. If a point is consistently identified as locally deep at these large locality levels, its frequency will be accordingly high. Conversely, if a point is only locally deep at small locality levels, its frequency will be low. Figure~\ref{fig:ILDexample} illustrates this behavior. Point $70$, which corresponds to a global depth median for the complete sample is consistently chosen as the locally deep point of most neighborhoods for both points, when the locality level is large enough. The frequency order thus reflects a natural ranking of representativeness among the defined locally deep points.
\subsection{Possible Application}
Representative points have been widely used in both supervised and unsupervised learning. Here, we explore a simple application of locally deep points by incorporating them into a $K$medoids clustering procedure, with details shown in Algorithm \ref{alg:depth_kmedoids}.

\begin{table*}[b]
\small
\caption[ARI values of $K$medoids variants across five datasets.
]{
ARI values of $K$medoids variants across five datasets.
The data set Olive is from the \texttt{pgmm} package~\cite{pgmm}.
Other datasets are discussed in Section~\ref{sec:rd}.
\textbf{Depth} and \textbf{Depth-2} correspond to Algorithm~\ref{alg:depth_kmedoids} with 
\texttt{only\_exemplars} set to \texttt{TRUE} and \texttt{FALSE}, respectively;
\textbf{PAM} refers to the standard method implemented in the \texttt{cluster} package~\cite{cluster}.
}
\label{tab:medoid}
\centering
\scalebox{1}[1]{
\begin{tabularx}{0.9\textwidth}{
>{\raggedright\arraybackslash}m{5.5cm} >{\raggedright\arraybackslash}m{2.5cm} >{\raggedright\arraybackslash}m{2.5cm} >{\raggedright\arraybackslash}m{2.5cm}
} 
\toprule  
\textbf{Dataset} & \textbf{Depth} & \textbf{Depth-2} & \textbf{PAM} \\
\midrule
BC \quad $(n=569, d=30, K=2)$    & $0.7488$ & $0.7488$ & $0.6079$ \\
Seed \quad $(n=210, d=7, K=3)$    & $0.7103$ & $0.7103$ & $0.7103$ \\
Wine \quad $(n=178, d=13, K=3)$    & $0.7137$ & $0.7137$ & $0.7411$ \\
Olive \quad $(n=572, d=8, K=3)$   & $0.7451$ & $0.7451$ & $0.7247$ \\
Seg \quad $(n=2086, d=18, K=7)$& $0.5203$ & $0.5178$ & $0.4550$\\
\bottomrule
\end{tabularx}
}
\end{table*}

Compared to the standard $K$medoids algorithm, the proposed method introduces two modifications. First, the initialization phase selects the most representative locally deep points—ranked by their frequency—while enforcing diversity by ensuring that no newly chosen exemplar lies within the empirical $\beta$-neighborhood $\vecN_{\vecx}^\beta$ of previously selected ones. This strategy encourages both representativeness and separation among initial medoids, and can be extended to other clustering methods requiring initialization. Second, during each iteration, medoid updates are guided by data depth: depending on a user-specified setting, the medoid for each cluster is selected either from all cluster members or restricted to locally deep exemplars. Table~\ref{tab:medoid} presents a comparison of clustering performance across methods. Notably, restricting medoid candidates to only the locally deep points yields comparable results to using all points, which supports the representativeness of these locally deep points.

\begin{algorithm}[t]
\footnotesize
\caption{Depth-based $K$medoids clustering with exemplar initialization}
\label{alg:depth_kmedoids}
\begin{algorithmic}[1] 
\Require Data matrix $\vecX$, number of clusters $K$, ordered exemplar set $\mathcal{E}$, depth-based neighborhood information, logical parameter $\texttt{only\_exemplars}$, max iterations.
\Ensure Cluster assignments $\mathcal{C}$, final medoids $\mathcal{M}$.
\State Set locality level $\beta = 1 / (1.5K)$.
\State Initialize center list $\mathcal{M} = [\ ]$ with first center as top exemplar: $\mathcal{M}_1 = \mathcal{E}_1$.
\While{length$(\mathcal{M}) < K$}
    \State Append to $\mathcal{M}$ the first remaining exemplar in $\mathcal{E}$ not in $\bigcup_{\vecx\in\mathcal{M}} \vecN^\beta_\vecx$.
\EndWhile
\State Assign each point to the closest medoid in $\mathcal{M}$, and obtain initial cluster assignment $\mathcal{C}_{\text{prev}}$.
\Repeat
    \For{each cluster $k = 1$ to $K$}
     \State Extract indices $\mathcal{I}_k$ of cluster $k$.
        \If{$\texttt{only\_exemplars}$}
            \State Compute depth of exemplars $\mathcal{E}$ w.r.t. cluster $k$.
            \State Set new medoid $\mathcal{M}_k = \arg\max_{e \in \mathcal{E}} D_k(e)$.
        \Else
            \State Compute depth of cluster points w.r.t. cluster $k$.
            \State Set $\mathcal{M}_k = \arg\max_{i \in \mathcal{I}_k} D_k(i)$.
        \EndIf
    \EndFor
    \State Reassign each point to the closest medoid in $\mathcal{M}$, and obtain new assignment $\mathcal{C}_{\text{new}}$.
    \State Set $\mathcal{C}_{\text{prev}} \gets \mathcal{C}_{\text{new}}$.
\Until{convergence or iteration limit reached}
\end{algorithmic}
\end{algorithm}

\subsection{Local centers for A-DLCC}\label{sec:slc}
Now, our target is to define local centers from those locally deep points, analogue to the filtering procedures in DLCC. We expect a local center should be deep in its own neighborhood, and with comparatively high frequency. In order to define local centers, we first introduce the following definition.
\begin{thm}[Self Centrality Level]
Let $\vecx$ be a locally deep point and let $\mathrm{ILD}^{\beta}(\vecx)$ denote its ILD value at locality level ${\beta} \in (0,1]$. Define the \emph{self centrality level set} of $\vecx$ as
\[
\mathcal{B}(\vecx) := \left\{ {\beta} \in (0,1] : \mathrm{ILD}^{\beta}(\vecx) = \max_{\vecz \in \vecN^{\beta}_{\vecx}} \mathrm{ILD}^{\beta}(\vecz) \right\},
\]
i.e., the set of locality levels at which $\vecx$ attains the maximal ILD value within its own $\beta$-neighborhood $\vecN^{\beta_b}_{\vecx}$. If $\mathcal{B}(\vecx)$ is non-empty, the \emph{self centrality level} of $\vecx$ is defined as
\[
B^*(\vecx) := \arg\max_{{\beta} \in \mathcal{B}(\vecx)} \mathrm{ILD}^\beta(\vecx).
\]
\end{thm}
$\mathcal{B}(\vecx)$ is the set of levels $\beta$ for which the point $\vecx$ has the largest ILD among all points in its neighborhood at level $\beta$. Then $B^*(\vecx)$ is the locality level corresponding to the largest depth value achieved across all levels in $\mathcal{B}(\vecx)$.
\begin{thm}[Local Center]
Let $\vecx$ be a locally deep point with a non-empty self centrality level set $\mathcal{B}(\vecx)$ and corresponding self centrality level $B^*(\vecx)$. Let $\vecN_{\vecx}^{B^*(\vecx)}$ denote the neighborhood of $\vecx$ under locality level $B^*(\vecx)$. We say that $\vecx$ is a \emph{local center} if it satisfies the following stability condition:
\[
\frac{1}{\left| \left\{ \vecz :\, \vecx \in \vecN_{\vecz}^{B^*(\vecx)} \right\} \right|} \sum_{\{\vecz :\, \vecx\in\vecN_{\vecz}^{B^*(\vecx)}\}} \mathbb{I} \left\{ 
\vecx = \arg\max_{\vecu \in \vecN_{\vecz}^{B^*(\vecx)}} \mathrm{ILD}^{B^*(\vecx)}(\vecu) \right\} > 0.5,
\]
where $\vecN_{\vecz}^{B^*(\vecx)}$ is the neighborhood of $\vecz$ under the same locality level $B^*$, $|\cdot|$ represents cardinality and $\mathbb{I}\{\cdot\}$ is the indicator function.  
\end{thm}
In summary, a local center is a locally deep point that has a well-defined self centrality level $B^*(\vecx)$ and is most frequently identified as the locally deepest point within the neighborhoods it belongs to at this level. 
\section{Automatic-depth based local center clustering} \label{sec:adlcc}
With defined local centers, the next target for A-DLCC is to group them. The A-DLCC algorithm has three parts: defining group similarities without parameters, determining the number of clusters, and constructing temporary clusters.
\subsection{Group-level local similarity}
For clarity of latter discussions, we use $\mathrm{DS}(\cdot,\cdot)$ to denote the depth-based similarity, where $\mathrm{DS}(\vecx_i,\vecx_j)=\bbS_{i,j}$ corresponds to the similarity directed from 
$\vecx_j$ to $\vecx_i$, i.e., $D(\vecx_j \mid \mathbf{X}_{R\vecx_i})$. While the depth-based similarity matrix may be asymmetric, in A-DLCC we follow DLCC and use the symmetrized version obtained by averaging $\bbS_{i,j}$ and $\bbS_{j,i}$, and, for notational simplicity, we continue to denote it by $\bbS$. This symmetrized similarity is then used to define the assignment of data points to local centers.

Assume the local centers are denoted by $\vecC = \{\vecc_1, \vecc_2, \ldots, \vecc_T\}$. 
Note that the indices of $\vecc$ differ from those of the original dataset. Each data point is assigned to the group associated with its most similar local center. Specifically, for each $t = 1, \ldots, T$, we define
\begin{equation}\label{eq:G}
    \vecG_t = \left\{ \vecx_u \in \vecX \,\middle|\, \vecc_t = \arg\max_{\vecc_i\in\vecC} \mathrm{DS}(\vecx_u, \vecc_i) \right\},
\end{equation}
the group of points in the data set whose most similar center is $\vecc_t$.
As a result, the groups $\{ \vecG_t \}$ form a partition of $\vecX$ into $T$ groups, and overlap-based measures cannot be used to quantify the similarity between different groups. Instead, inspired by the concept of PaLD \cite{berenhaut2022social},  we design a similarity measure for disjoint groups. Specifically, for each pair of groups $(\vecG_i, \vecG_j)$, we define the group-level local similarity $\mathrm{GLS}(i, j)$ as follows.
For any randomly chosen $\vecx_u \in \vecG_i$ and $\vecx_v \in \vecG_j$, define the \emph{local focus region} as
\[
    U_{\vecx_u,\vecx_v} = \left\{ \vecz \in \vecX \;\middle|\; \mathrm{DS}(\vecx_u,\vecz) \geq \mathrm{DS}(\vecx_u,\vecx_v) \;\text{or}\; \mathrm{DS}(\vecx_v,\vecz) \geq \mathrm{DS}(\vecx_v,\vecx_u) \right\}.
\]
Define the directional similarity from $\vecG_j$ to $\vecG_i$ as
\[
W_{\vecG_j \to \vecG_i} = \frac{1}{|\vecG_i|\,|\vecG_j|} 
\sum_{\vecx_u \in \vecG_i} \sum_{\vecx_v \in \vecG_j} 
H\bigl(\vecx_u, \vecx_v \mid \vecG_j \cap U_{\vecx_u,\vecx_v}\bigr),
\]
where for any set $\vecG$,
\[
H(\vecx_u, \vecx_v \mid \vecG) = \frac{1}{|\vecG|} \sum_{\vecz \in \vecG} 
\left[
    \mathbb{I}\bigl(\mathrm{DS}(\vecz, \vecx_u) > \mathrm{DS}(\vecz, \vecx_v)\bigr)
    + \frac{1}{2}\, \mathbb{I}\bigl(\mathrm{DS}(\vecz, \vecx_u) = \mathrm{DS}(\vecz, \vecx_v)\bigr)
\right].
\]
Then, the \emph{group-level local similarity} between $\vecG_i$ and $\vecG_j$ is the symmetric average
\[
\mathrm{GLS}(i, j) = \frac{1}{2} \left( W_{\vecG_i \to \vecG_j} + W_{\vecG_j \to \vecG_i} \right).
\]
Based on this, we define $\mathbb{G} = \left\{ \mathrm{GLS}(i,j) \right\}_{i,j \in \{1, \ldots, T\}}
$ as the group-level local similarity matrix.

We note that similarity \(W_{\vecG_j \to \vecG_i}\) admits the following probabilistic interpretation. Randomly sample a pair of points \(V_i \in \vecG_i\), \(V_j \in \vecG_j\). Then, uniformly sample a point \(Z \in U_{V_i,V_j} \cap \vecG_j\). The value \(W_{\vecG_j \to \vecG_i}\) represents the probability that such a point \(Z\) is more similar to \(V_i \in \vecG_i\) than to \(V_j \in \vecG_j\), i.e.,
\[
W_{\vecG_j \to \vecG_i} = \mathbb{P} \left(\mathrm{DS}(Z,V_i)>\mathrm{DS}(Z,V_j) \right).
\]
Ties, if any, are broken uniformly at random. A large value of \(W_{\vecG_j \to \vecG_i}\) implies that \(\vecG_j\) is not well-separated from \(\vecG_i\).
The GLS matrix symmetrizes this relationship by averaging \(W_{\vecG_j \to \vecG_i}\) and \(W_{\vecG_i \to \vecG_j}\), giving a balanced estimate of the mutual similarity between the two groups.
\begin{pro}[Zero similarity under strong separation] \label{pro:zs}
Let $\vecG_i, \vecG_j \subseteq \vecX$ be two disjoint groups. Suppose that, for all $\vecx_u \in \vecG_j$, we have
\[
\min_{\vecx_v \in \vecG_j}  \mathrm{DS}(\vecx_u, \vecx_v) > \max_{\vecz \in \vecG_i}  \mathrm{DS}(\vecx_u, \vecz).
\]
Then the directional group similarity from $\vecG_j$ to $\vecG_i$ satisfies $
W_{\vecG_j \to \vecG_i} = 0.
$
\end{pro}
In practice, only similarities between nearby groups are typically of interest. To accelerate processing, one can omit GLS evaluations for group pairs that satisfy the following centroid-based separation check
$$
\min_{\vecx_u, \vecx_v \in \vecG_j} \mathrm{DS}(\vecx_u, \vecx_v) > \max_{\vecz \in \vecG_i} \mathrm{DS}(\vecc_j, \vecz), \qquad \min_{\vecx_u, \vecx_v \in \vecG_i} \mathrm{DS}(\vecx_u, \vecx_v) > \max_{\vecz \in \vecG_j} \mathrm{DS}(\vecc_i, \vecz).
$$
Although this condition does not guarantee that $\mathrm{GLS}(i, j) = 0$, the similarity is generally small enough that omitting the computation should have negligible impact on the subsequent adaptive grouping processes.
\begin{pro}[Invariance under similarity transformations]\label{pro2:inva}
For any pair of groups $\vecG_i$ and $\vecG_j$, if a similarity transformation $T$ (i.e., any combination of rotation, dilation, and translation) is applied to $\vecG_i\cup\vecG_j$, then $\mathrm{GLS}(i, j)$ remains unchanged for any input similarity measure $\mathrm{DS}$ that preserves the ordinal relationships among points under the transformation, i.e., if $\mathrm{DS}(\vecx_u,\vecx_v)>\mathrm{DS}(\vecx_u,\vecx_q)$, then $\mathrm{DS}(T(\vecx_u),T(\vecx_v))>\mathrm{DS}(T(\vecx_u),T(\vecx_q))$.
\end{pro}
In particular, the invariance of $\bbS$ under similarity transformations ensures that Proposition~\ref{pro2:inva} holds directly in our setting, that is, $\vecG$ is also invariant under similarity transformation.
\subsection{Adaptive grouping strategy}\label{sec:ags}
The group similarity alone does not determine whether two groups should merge. We introduce an adaptive grouping rule that decides, for every pair of adjacent groups, whether one is part of the other (absorption) or whether the two are two parts of one structure (bonding). The rule builds on the bottleneck path \cite{gabow1988algorithms,chebotarev2011graph} concept in graph theory. The idea is to measure the strength of connection between two nodes as the strongest path between them, where each path’s strength is defined by its weakest edge. We adopt this idea to define reachable similarity using symmetric similarity matrices. 
\begin{thm}[Reachable similarity]\label{def:rs}
    Let $N$ be the number of nodes, and let $S$ be a symmetric similarity matrix with entries $S_{i,j}$ in $[0,1]$. The \emph{reachable similarity} $\mathrm{rs}(i, j)$ between any two nodes $i$ and $j$ is defined as the maximum, over all paths connecting $i$ to $j$, of the minimum similarity along that path. Formally, for any path $\boldsymbol{p} = (p_0, p_1, \dots, p_m)$ with $p_0 = i$ and $p_m = j$, its bottleneck similarity is
    $
    \min_{0 \leq \ell < m} S_{p_\ell, p_{\ell+1}}.
    $
    Then, the reachable similarity between nodes $i$ and $j$ is given by
    \[
    \mathrm{rs}(i,j) = \max_{\boldsymbol{p}\,:\,i \to j} \; \min_{0 \leq \ell < m} S_{p_\ell, p_{\ell+1}}.
    \]
\end{thm}
In our framework, $S$ may denote either the pointwise similarity matrix $\bbS$ or the groupwise similarity matrix $\mathbb{G}$. We denote their corresponding reachable similarity matrices as $\tilde{\bbS}$ and $\tilde{\mathbb{G}}$, respectively. 

\begin{thm}[Intra- and between-group reachability]
For any groups $\vecG_i$ and $\vecG_j$, and for any $\vecx \in \vecG_i$, define
\begin{align}
    \text{Intra-group similarity:}\quad & \mu_{\vecG_i}(\vecx) = \frac{1}{|\vecG_i| - 1} \sum_{\substack{\vecz \in \vecG_i\setminus\vecx}} \mathrm{rs}_{\vecG_i}(\vecx, \vecz), \\
    \text{Between-group similarity:}\quad & \mu_{\vecG_i \to \vecG_j}(\vecx) = \frac{1}{|\vecG_j|} \sum_{\vecz \in \vecG_j} \mathrm{rs}_{\vecG_i \cup \vecG_j}(\vecx, \vecz),
\end{align}
where $\mathrm{rs}_{\vecG}(\cdot, \cdot)$ denotes the reachable similarity computed within the set $\vecG$, and $|\vecG_i|$ and $|\vecG_j|$ are the sizes of $\vecG_i$ and $\vecG_j$, respectively.
\end{thm}
\begin{thm}[Relative reachability ratio] \label{def:rrr}
Given two groups $\vecG_i$ and $\vecG_j$, the relative reachability ratio from $\vecG_i$ to $\vecG_j$ is defined as
\begin{equation}
    \rho_{\vecG_i \to \vecG_j} = \frac{1}{|\vecG_i|} \sum_{\vecx \in \vecG_i}  \frac{\mu_{\vecG_i \to \vecG_j}(\vecx)}{\mu_{\vecG_i}(\vecx)},
\end{equation}
and the quantity
$
\min\{\rho_{\vecG_i \to \vecG_j} , \rho_{\vecG_j \to \vecG_i}\}
$
is referred to as the \emph{minimal reachability ratio} of the merge between $\vecG_i$ and $\vecG_j$.
\end{thm}
We now introduce the acceptance rule for determining whether two groups should be merged. Different clustering scenarios may require different minimal reachability ratios. For connected shapes, stricter thresholds help avoid incorrect merges, while in well-separated settings, lower thresholds are acceptable. We therefore assign a base threshold to each group by evaluating the relative reachability ratio $\rho_{\vecG_i \to \vecX\setminus\vecG_i}$,which we call the \emph{background level} of $\vecG_i$ and denote by $\bar\phi_i=\rho_{\vecG_i \to \vecX\setminus\vecG_i}$. The base threshold is its clipped version
\begin{equation}
    \phi_i=  \min\left( \max(\bar\phi_i,\, 0.9),\, 0.99 \right). 
\end{equation}
The bounding interval $[0.9, 0.99]$ ensures that the threshold remains within a reasonable range, avoiding values that are too permissive or overly strict. A background level $\bar\phi_i\ge1$ means that the points of $\vecG_i$ reach the rest of the data at least as well as they reach each other. This case is treated separately in Section~\ref{sec:cells}.

This base threshold prevents all groups from collapsing into one, but pairwise merges usually need a stricter criterion. We define a merge-specific acceptance threshold that penalizes merges with large disruption. Consider two groups $\vecG_i$ and $\vecG_j$, and let $\mu_{\vecG_{ij}}(\vecx)=\mu_{\vecG_i \cup \vecG_j}(\vecx)$ denote the average reachable similarity of $\vecx$ to all other points in $\vecG_i \cup \vecG_j$ (i.e., its intra-group reachability after merging). We define the \emph{disruption} $\omega_{i|j}$ suffered by $\vecG_i$ in the merge as the proportion of its points whose intra-group reachability decreases after merging,
\begin{equation} \label{eq:omega}
\omega_{i|j} = 
    \frac{|\{\vecx \in \vecG_i : \mu_{\vecG_i}(\vecx) > \mu_{\vecG_{ij}}(\vecx)\}|}{|\vecG_i|},
\end{equation}
and $\omega_{j|i}$ symmetrically. Keeping the two sides separate, rather than averaging them, lets each side judge the merge from its own point of view; a small group swallowed by a large one and a large group touched by a small one are disrupted very differently. The adaptive threshold that $\vecG_i$ applies to the merge with $\vecG_j$ is
\begin{equation}\label{eq:adapth}
Q_{\mathrm{th}}(i\mid j) = \phi_i + \omega_{i|j}^2\,\bigl(\max(q,\phi_i) - \phi_i\bigr),\qquad q=0.975.
\end{equation}
Here, the effect of $\omega$ is dampened by squaring, to avoid overreacting to mild degradations that commonly occur when merging distinct groups. The threshold interpolates between the background level of $\vecG_i$ and a fixed upper level $q$. In all experiments $q$ is fixed as $0.975$ without tuning.

\subsubsection{Isolated local centers and cells without a boundary}\label{sec:cells}
Two checks precede the merging rule.

The first check covers small and simple data sets in which each cluster produces a single local center. If the local centers are far from each other relative to the size of their groups, no merge should even be considered. Given locality level $\beta=\max_t|\vecG_t|/2n$, we check the neighbors of each local center: if no other local center appears in the neighborhood $\vecN_{\vecc_t}^{\beta}$ of any local center $\vecc_t$, all local centers are treated as singleton groups and the merging step is skipped altogether.

The second check identifies groups that have no boundary of their own. When $\bar\phi_i=\rho_{\vecG_i\to\vecX\setminus\vecG_i}\ge 1$, the points of $\vecG_i$ reach the rest of the data at least as well as they reach each other, so $\vecG_i$ is only a candidate. A candidate is set aside if its mean intra-group reachability is below that of every non-candidate linked to it by a direct GLS edge ($\mathbb{G}_{i,j}=\tbbG_{i,j}\neq 0$ and $\bar\phi_j<1$); if it has no such neighbor, it is set aside only when its size is smaller than the median group size. Set-aside groups skip the merging rule, and their local centers are demoted to ordinary points. We call the remaining groups \emph{units}. After the units have been grouped, every observation is assigned to the retained local center it is most similar to, so the points originally in a set-aside group join a unit.

\subsubsection{Absorption and bonding}\label{sec:rule}
Let $\vecg_1,\ldots,\vecg_K$ be the current groups of local centers and for a group $\vecg$, let $\vecG_{\vecg}=\bigcup_{t:\vecc_t\in\vecg}\vecG_t$ be its points. The matrices $\mathbb{G}$ and $\tbbG$ are not updated as these groups are merged. An entry $\mathbb{G}_{i,j}$ remains the group-level similarity between the initial groups of local centers $\vecc_i$ and $\vecc_j$. Two groups are \emph{adjacent} when some pair of their local centers has positive group-level local similarity. For every adjacent pair $(\vecg_a,\vecg_b)$ we compute the two relative reachability ratios $\rho_{a\to b}=\rho_{\vecG_{\vecg_a}\to\vecG_{\vecg_b}}$ and $\rho_{b\to a}$ and the two disruptions $\omega_{a|b}$ and $\omega_{b|a}$, and admit the pair to exactly one of two kinds of merge. The ratios themselves decide which one. If at least one of $\rho_{a\to b}$ and $\rho_{b\to a}$ is $\ge 1$, that side reaches the other at least as well as it reaches itself and has no boundary of its own relative to the other; the two groups are not peers, and the pair is examined for absorption only. If both ratios are below $1$, each group is self-contained relative to the other; the two are peers, and the pair is examined for bonding only.
\begin{enumerate}
    \item \textbf{Absorption.}
 Group $\vecg_s$ is absorbed by group $\vecg_w$ when
\begin{equation}\label{eq:absorb}
    \rho_{s\to w}\ \ge\ 1
    \qquad\text{and}\qquad
    \rho_{w\to s}\ >\ \min\bigl(\bar\phi_{w}^{-s},\,0.99\bigr),
\end{equation}
where
\begin{equation}\label{eq:philoo}
    \bar\phi_{w}^{-s}=\frac{1}{|\vecG_{\vecg_w}|}\sum_{\vecx\in\vecG_{\vecg_w}}\frac{1}{\mu_{\vecG_{\vecg_w}}(\vecx)}\cdot\frac{1}{|\mathcal{B}|}\sum_{\vecz\in\mathcal{B}}\mathrm{rs}_{\vecX}(\vecx,\vecz),
    \qquad \mathcal{B}=\vecX\setminus(\vecG_{\vecg_w}\cup\vecG_{\vecg_s}),
\end{equation}
is the background level of $\vecg_w$ after $\vecg_s$ has been left out of the reference set $\mathcal{B}$. If $\vecg_s$ is large and nearby it lifts $\bar\phi_w$, so $\bar\phi_w^{-s}$ averages only over $\mathcal{B}$. The reachable similarities are the same $\mathrm{rs}_{\vecX}$ used for $\bar\phi_w$: they come from the full data, and $\bar\phi_w^{-s}$ differs from $\bar\phi_w$ only in which points enter the outer average. $\rho_{s\to w}\ge 1$ means $\vecg_s$ reaches $\vecg_w$ at least as well as it reaches itself, hence has no boundary of its own relative to $\vecg_w$. $\rho_{w\to s}>\min(\bar\phi_w^{-s},0.99)$ means $\vecg_w$ reaches $\vecg_s$ more than it reaches the rest of the data with $\vecg_s$ left out; the $0.99$ is the same cap already used for $\phi_i$. A group that could be absorbed by several neighbors goes with the one maximizing $\rho_{s\to w}$.

\item \textbf{Bonding.} Two peers, i.e., groups with $\rho_{a\to b}<1$ and $\rho_{b\to a}<1$, bond when each side finds the other more reachable than its own threshold,
\begin{equation}\label{eq:bond}
    \rho_{a\to b}\ >\ Q_{\mathrm{th}}(a\mid b)
    \qquad\text{and}\qquad
    \rho_{b\to a}\ >\ Q_{\mathrm{th}}(b\mid a),
\end{equation}
with $Q_{\mathrm{th}}$ as in~\eqref{eq:adapth}. Once either group has internal links, bonding is subject to one further condition, that it creates no new weakest link. Let $\mathrm{b}(\vecg)$ be the weakest edge of the maximum spanning tree of $\mathbb{G}$ restricted to the local centers of $\vecg$, i.e., the weakest group-level similarity on which the internal connectivity of $\vecg$ already relies ($\mathrm{b}(\vecg)=\infty$ for a singleton). When at least one of the two bottlenecks is finite, the bond is accepted only if their contact is not weaker than what either of them already tolerates,
\begin{equation}\label{eq:dip}
    \max_{\vecc_i\in\vecg_a,\,\vecc_j\in\vecg_b}\mathbb{G}_{i,j}\ \ge\ q\,\min\{\mathrm{b}(\vecg_a),\mathrm{b}(\vecg_b)\},
\end{equation}
with the same slack $q$ as in~\eqref{eq:adapth}. When both groups are singletons, both bottlenecks are infinite and~\eqref{eq:dip} is not imposed. Condition~\eqref{eq:dip} is what keeps elongated structures from being chained through a dip in similarity while letting fragments of one structure, whose mutual contacts are as strong as their internal ones, be reunited.
\end{enumerate}
\subsubsection{Community-level contact}\label{sec:null}
Conditions~\eqref{eq:absorb} and~\eqref{eq:bond} compare two groups with each other and with their own background, but they do not account for the size of the groups relative to the whole data set. The same reachability ratios are more likely to reflect genuine structure when the two groups are small, whereas for two large groups the same ratios can arise simply because their size makes substantial contact statistically expected; merging in the latter case carries a much higher risk. To correct for this size effect we compare the observed contact between two groups with the contact expected from their size alone, using a standard null model of graph community structure.

Let $\mathcal{D}$ be the \emph{depth graph} on $\vecX$: every observation $\vecx_u$ is linked to its $r_u$ most similar observations, where $r_u=|\vecG_t|$ for the group $\vecG_t$ containing $\vecx_u$, the edge weights are the depth-based similarities stored in $\bbS$, and the graph is symmetrized by taking, for each pair, the larger of the two directed weights. For two groups $\vecG_i$ and $\vecG_j$ let $e_{ij}$ be the total weight of the edges between them, $k_i$ and $k_j$ the total weights incident to them, and $m$ the total weight of $\mathcal{D}$. The modularity gain of joining the two groups~\cite{newman2004finding} is
\begin{equation}\label{eq:dq}
    \Delta Q_{ij}=\frac{e_{ij}}{m}-\frac{k_i\,k_j}{2m^2}.
\end{equation}
$\Delta Q_{ij}>0$ means that the two groups share more weight than the configuration model, which places edges at random given the weight of every observation, would predict for two groups of their size. We call such a pair a \emph{community-level contact}. A merge of $\vecg_a$ and $\vecg_b$, whether absorption or bond, is accepted only if at least one pair of groups $\vecG_i\subseteq\vecG_{\vecg_a}$, $\vecG_j\subseteq\vecG_{\vecg_b}$ is a community-level contact. The test is carried out at the level of the initial groups $\vecG_t$ and not at the level of $\vecg_a$ and $\vecg_b$ as wholes, because the modularity of a partition has a resolution limit~\cite{fortunato2007resolution}: the gain of joining two large communities is negative even when they are genuinely connected, whereas the gain between two adjacent initial groups is not affected by the size of the groups they have grown into. One kind of absorption is exempt from the test. When $\omega_{s|w}=0$, no point of $\vecg_s$ loses intra-group reachability in the union, so $\vecg_s$ is not a community adjacent to $\vecg_w$ but a part of its reachability basin; the question the null model asks does not arise.

\subsubsection{Iteration and reconsideration of small groups}\label{sec:iter}
The rule is applied to all adjacent pairs of the current groups at once; the new groups are the connected components of the accepted absorptions and bonds. The pass is repeated on the new groups, with $\vecG_{\vecg}$, $\bar\phi$, $\omega$ and $\mathrm{b}(\cdot)$ recomputed for the merged groups, until no merge is accepted. Nothing changes between the first pass over singletons and the later passes over groups except that condition~\eqref{eq:dip} only becomes active once a group has internal links.

Finally, small groups are reconsidered. For each group $\vecg_k$ let $\mathcal{T}_k=\bigcup_{t:\vecc_t\in\vecg_k}\vecG_t$ be its temporary cluster, where the $\vecG_t$ are recomputed over the retained local centers only, so that the observations of groups that were set aside in Section~\ref{sec:cells} are included. Every $\mathcal{T}_i$ with fewer points than half the median cluster size is paired with the cluster $\mathcal{T}_j$ it is most similar to, where the similarity between two groups of local centers is $\max\tbbG_{u,v}\,\mathbb{G}_{u,v}$ over their local centers, balancing direct and reachable similarity. Denoting the union by $\mathcal{T}_{ij}=\mathcal{T}_i\cup\mathcal{T}_j$, the merge is accepted if the union is not more connected to the outside than the looser of the two parts already was,
\begin{eqnarray}\label{eq:rhocompare}
    \rho_{\mathcal{T}_{ij}\to\vecX\setminus\mathcal{T}_{ij}}\ \le\ \max\bigl\{\rho_{\mathcal{T}_{i}\to\vecX\setminus\mathcal{T}_{i}},\ \rho_{\mathcal{T}_{j}\to\vecX\setminus\mathcal{T}_{j}}\bigr\},
\end{eqnarray}
provided the union is not larger than the largest current cluster and the pair contains a community-level contact. A small group that passes~\eqref{eq:rhocompare} is a fragment whose points are better described as part of a neighbor than as a mode of their own; a small group that fails it is kept, however small. The step is repeated until no small group moves. The complete procedure is summarized in Algorithm~\ref{alg:reach}.

\begin{algorithm}[t]
\footnotesize
\caption{Adaptive grouping of local centers}
\label{alg:reach}
\begin{algorithmic}[1]
\Require Local centers $\vecC=\{\vecc_1,\ldots,\vecc_T\}$, point groups $\vecG_1,\ldots,\vecG_T$, similarity matrix $\bbS$, group-level similarity matrix $\mathbb{G}$ and its reachable version $\tbbG$, slack $q=0.975$.
\Ensure Groups of local centers $\vecg_1,\ldots,\vecg_K$ and temporary clusters $\mathcal{T}_1,\ldots,\mathcal{T}_K$.
\State If no local center lies in the $\beta$-neighborhood of another one with $\beta=\max_t|\vecG_t|/2n$, return the singletons.
\State Compute $\bar\phi_t$ for all groups; set aside the groups with $\bar\phi_t\ge1$ that meet the criteria of Section~\ref{sec:cells}; the remaining groups are the units.
\State Build the depth graph $\mathcal{D}$ and compute $\Delta Q_{ij}$ by~\eqref{eq:dq} for all adjacent pairs of units.
\State Initialize $\vecg_1,\ldots,\vecg_K$ as the singletons of the units.
\Repeat
    \State For every adjacent pair $(\vecg_a,\vecg_b)$ compute $\rho_{a\to b},\rho_{b\to a}$ and $\omega_{a|b},\omega_{b|a}$ on $\vecG_{\vecg_a}$ and $\vecG_{\vecg_b}$.
    \State Among the pairs with $\max(\rho_{a\to b},\rho_{b\to a})\ge1$, mark as absorptions those satisfying~\eqref{eq:absorb}, each absorbed group keeping only the neighbor it reaches best; the other pairs of this kind are left unmerged.
    \State Among the pairs with $\max(\rho_{a\to b},\rho_{b\to a})<1$, mark as bonds those satisfying~\eqref{eq:bond} and, when either group has internal links,~\eqref{eq:dip}.
    \State Discard marked pairs without a community-level contact, unless the pair is an absorption with $\omega_{s|w}=0$.
    \State Replace the groups by the connected components of the marked pairs; recompute $\bar\phi$ and $\mathrm{b}(\cdot)$ for the new groups.
\Until{no pair is marked}
\State Assign every observation to its most similar retained local center; form $\mathcal{T}_k=\bigcup_{t:\vecc_t\in\vecg_k}\vecG_t$.
\While{some $|\mathcal{T}_i|<\frac12\,\mathrm{median}(|\mathcal{T}_1|,\ldots,|\mathcal{T}_K|)$ has not been examined}
    \State Pair $\mathcal{T}_i$ with the cluster $\mathcal{T}_j$ maximizing $\max_{\vecc_u\in\vecg_i,\vecc_v\in\vecg_j}\tbbG_{u,v}\mathbb{G}_{u,v}$.
    \State Merge $\vecg_i$ into $\vecg_j$ if~\eqref{eq:rhocompare} holds, $|\mathcal{T}_{ij}|\le\max_k|\mathcal{T}_k|$ and the pair contains a community-level contact; recompute the cluster sizes.
\EndWhile
\end{algorithmic}
\end{algorithm}

The rule has no strategy switch and no data-set-specific branch. Absorption is what the ``min'' strategy of DLCC does when it attaches a non-representative local center to a representative one, and bonding is what the ``max'' strategy does when it links fragments of a connected shape; the two are decided pair by pair from the same quantities, so a data set can contain both situations at once. Figure~\ref{fig:temproray} in Section~\ref{sec:tc} illustrates the resulting groups on two toy examples, a data set of interlocked shapes and one of eleven normal components, in which the two kinds of merge dominate respectively.

\subsection{Temporary clusters}\label{sec:tc}
In the DLCC framework, the next step is to construct temporary clusters by partitioning the data into two parts: observations that can be confidently clustered, and those left unlabeled. The former are grouped into what we term temporary clusters. Unlike the original DLCC, which relies on unique neighbors based on a fixed locality parameter, here we define retained points, a parameter-free alternative, to construct temporary clusters.
\begin{thm}[Retained Points]\label{def:rp}
Let $\vecg_1, \ldots, \vecg_K$ denote $K$ groups of local centers, and let $\vecG_t$ denote the group of points assigned to local center $\vecc_t$ as defined in~\eqref{eq:G}, for $t = 1,\ldots,T$. Define the initial temporary cluster as $\mathcal{T}_k = \bigcup_{t:\vecc_t \in \vecg_k} \vecG_t$. A point $\vecx_p \in \vecG_t \subset \mathcal{T}_k$ is called a \emph{retained point} of $\mathcal{T}_k$ if
$$ 
\max_{\vecx_q \in \mathcal{T}_k \setminus \vecG_t} \mathrm{DS}(\vecx_p,\vecx_q) > \max_{\vecx_h \in \bigcup_{j \ne k} \mathcal{T}_j}  \mathrm{DS}(\vecx_p,\vecx_h),
$$
meaning that the maximum similarity between $\vecx_p$ and any point in $\mathcal{T}_k \setminus \vecG_t$ exceeds its maximum similarity with points in other temporary clusters.
\end{thm}

Definition~\ref{def:rp} assumes that $\vecg_k$ contains multiple local centers, which may not always be the case. We handle singleton groups $\vecg_k = \{\vecc_u\}$ as follows. Recall that $\mathcal{E}' = \mathcal{E} \setminus \vecC$, as defined in Remark~\ref{rem:E}, consists of the locally deep points not selected as local centers. Consider points $\vecz \in \mathcal{E}'$ that are more similar to $\vecc_u$ than to any other local center, i.e., $\mathrm{DS}(\vecz,\vecc_u) > \max_{\vecc_t \in \vecC \setminus {\vecc_u}} \mathrm{DS}(\vecz,\vecc_t)$. For each such point, we compute its depth with respect to $\vecG_u$ and weight it by a frequency-based coefficient
\begin{equation}\label{eq:fwDepth}
D_{f}(\vecz) =\frac{|\mathcal{E}'| - \sum_{\vecx \in \mathcal{E}'} \mathbb{I} \left[ f_{\vecx} > f_{\vecz} \right]}{|\mathcal{E}'|} D(\vecz \mid \vecG_u) ,
\end{equation}
where $f_{\vecx}$ is the frequency of $\vecx$. The frequency-based weight favors candidates that are more frequently identified as the \emph{locally deep point} across varying neighborhoods, while the depth value quantifies points' centrality within $\vecG_u$. A high depth value helps reduce the risk of including a locally deep point that is ambiguous or potentially misassigned to this group. The point $\vecz$ maximizing $D_f$ then serves as a local center and is added to the singleton group  $\vecg_k$ to ensure that Definition~\ref{def:rp} is well-defined.

The remaining steps follow the temporary cluster update procedure of the ``min strategy'' from the original DLCC algorithm~\cite{wang2025depth} with minor adaptations to accommodate the nonparametric setting, using the same score function. The score for assigning a point $\vecx_i$ to the $k$th temporary cluster is defined as
\begin{eqnarray}\label{eq:scoremin}
\operatorname{score}_{i|k}=\frac{\max_{\vecc_j \in \vecg_k} \mathrm{DS}(\vecc_j,\vecx_i)- \max_{\vecc_t \in \vecC\setminus\vecg_k} \mathrm{DS}(\vecc_t,\vecx_i)}{\max\{\max_{\vecc_j \in \vecg_k} \mathrm{DS}(\vecc_j,\vecx_i) , \max_{\vecc_t \in \vecC\setminus\vecg_k} \mathrm{DS}(\vecc_t,\vecx_i)\} \,}.
\end{eqnarray}
 The full procedure is summarized in Algorithm~\ref{alg:utc}, where, instead of relying on a predefined neighborhood size parameter, the output temporary cluster $\mathcal{T}_k$ is required to retain at least half of its initial size.
\begin{algorithm}[b!]
\footnotesize
\caption{Generate and update temporary clusters}
\label{alg:utc}
\begin{algorithmic}[1] 
\Require Groups of local centers $\{\vecg_1, \ldots, \vecg_K\}$, groups of points $\vecG_1,\ldots,\vecG_T$, similarity matrix $\bbS$, locally deep points $\mathcal{E}'$.
\Ensure Temporary clusters $\mathcal{T}_1,\ldots,\mathcal{T}_K$
\State For any singleton $\vecg_k = \{\vecc_u\}$, add a locally deep point $\vecz$ with the highest $D_f(\vecz\mid\vecG_s)$ value; update $T$, and current groups of points $\vecG_1,\ldots,\vecG_T$.
\State Initialize $\mathcal{T}_k \gets \bigcup_{t:\vecc_t \in \vecg_k} \vecG_t$ for $k = 1,\ldots,K$
\State Update each $\mathcal{T}_k$ using Definition~\ref{def:rp}, let remaining points be $\hat{\mathcal{T}}_k$.
    \State Generate score pools $\mathcal{S}_k$ and $\hat{\mathcal{S}}_k$ for points in each $\mathcal{T}_k$ and $\hat{\mathcal{T}}_k$ by \eqref{eq:scoremin}.
    \State Set $\alpha \gets \begin{cases}
    \mathrm{mean}(\bigcup_{k=1}^K \hat{\mathcal{S}}_k), & \text{if } \bigcup_{k=1}^K \hat{\mathcal{S}}_k \neq \emptyset \\
    0, & \text{otherwise}
\end{cases}$   
    \For{$k = 1$ to $K$}
      \State $\texttt{size}\gets|\mathcal{S}_k|+|\hat{\mathcal{S}}_k|$
       \If{$|\hat{\mathcal{S}}_k| > |\mathcal{S}_k|$}
        \State Set $\hat{\alpha}_1 \gets 0.5\,\alpha$
    \Else
        \State Set $\hat{\alpha}_1 \gets \alpha$
    \EndIf
        \For{each $i \in \mathcal{T}_k$ with score $< \hat{\alpha}_1$}
            \State $\mathcal{T}_k \gets \mathcal{T}_k \setminus \{i\}$,\quad $\hat{\mathcal{T}}_k \gets \hat{\mathcal{T}}_k \cup \{i\}$, and update $\mathcal{S}_k$ and $\hat{\mathcal{S}}_k$ accordingly.
        \EndFor
        \State Let $SS \gets$ sorted scores in $\mathcal{S}_k$ (descending)
      \State $idx \gets \argmax_{j = \text{floor}(|SS|/2) \text{ to } |SS|-1} (SS[j] - SS[j+1])$    
    \State $\hat{\alpha}_2 \gets \min(SS[idx],  1 - \frac{\texttt{size}/2 - |SS|}{|\hat{\mathcal{S}}_k|} \text{quantile of } \hat{\mathcal{S}}_k)$
        \For{each $u \in \hat{\mathcal{T}}_k$ with score $> \hat{\alpha}_2$}
        \State $\hat{\mathcal{T}}_k \gets \hat{\mathcal{T}}_k \setminus \{u\}$,\quad $\mathcal{T}_k \gets \mathcal{T}_k \cup \{u\}$.
        \EndFor
    \EndFor
\end{algorithmic}
\end{algorithm}

\begin{figure}[ht]
    \centering
    \begin{subfigure}[b]{0.45\textwidth}
        \includegraphics[width=\linewidth]{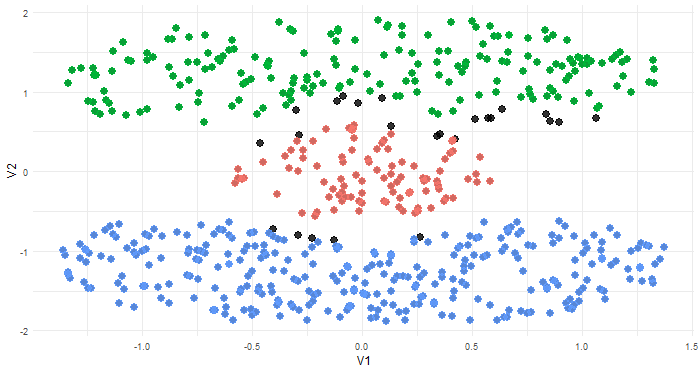}
        \caption{}
        \label{fig:tb1}
    \end{subfigure}
    \hfill
        \begin{subfigure}[b]{0.45\textwidth}
        \includegraphics[width=\linewidth]{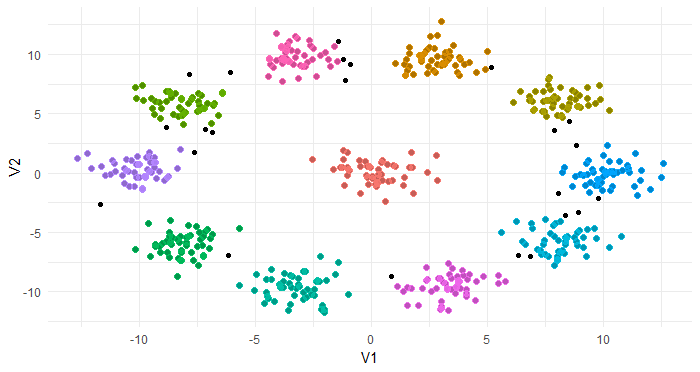}
        \caption{}
        \label{fig:tn2}
    \end{subfigure}
    \caption{Temporary cluster results for two toy examples. 
(a) Bainba, where the groups are formed by bonding fragments of the interlocked shapes; (b) Normal, eleven normal components, where the groups are formed by absorption of the less representative local centers.
Colors indicate retained points in each temporary cluster, while black points are unlabelled and left to be classified.}
    \label{fig:temproray}
\end{figure}

Figure~\ref{fig:temproray} illustrates the temporary clusters in the two toy examples. Most points are correctly clustered, while ambiguous points—typically those near cluster boundaries—remain unlabelled. 
The remaining points are then classified using classical methods.
\section{Applications} \label{sec:app}
We first compare A-DLCC with DLCC on several synthetic datasets, then with other methods considered ``adaptive'' or ``automatic'' on real datasets. 
\subsection{Synthetic data}
\begin{figure}[htbp]
    \centering
    \begin{subfigure}[b]{0.23\textwidth}
        \includegraphics[width=\linewidth]{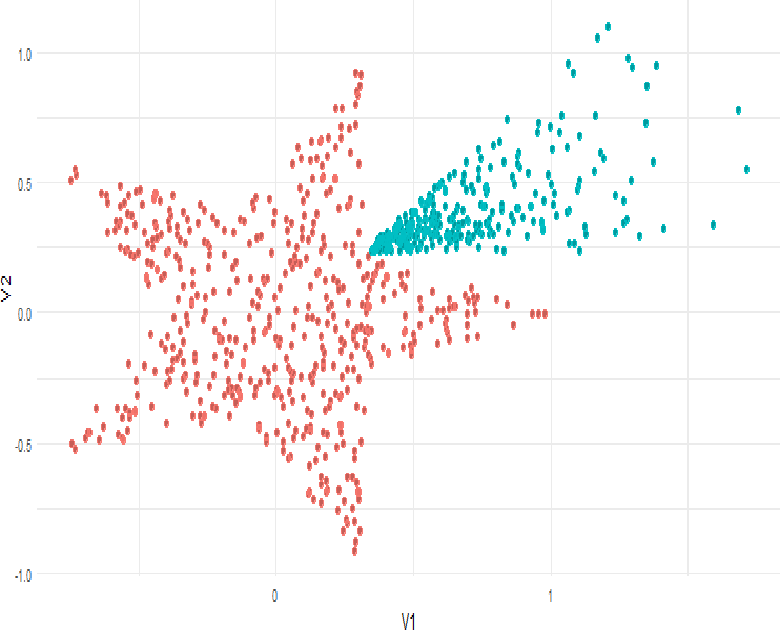}
        \caption{DLCC (Starbeam)}
    \end{subfigure}
    \hfill
    \begin{subfigure}[b]{0.23\textwidth}
        \includegraphics[width=\linewidth]{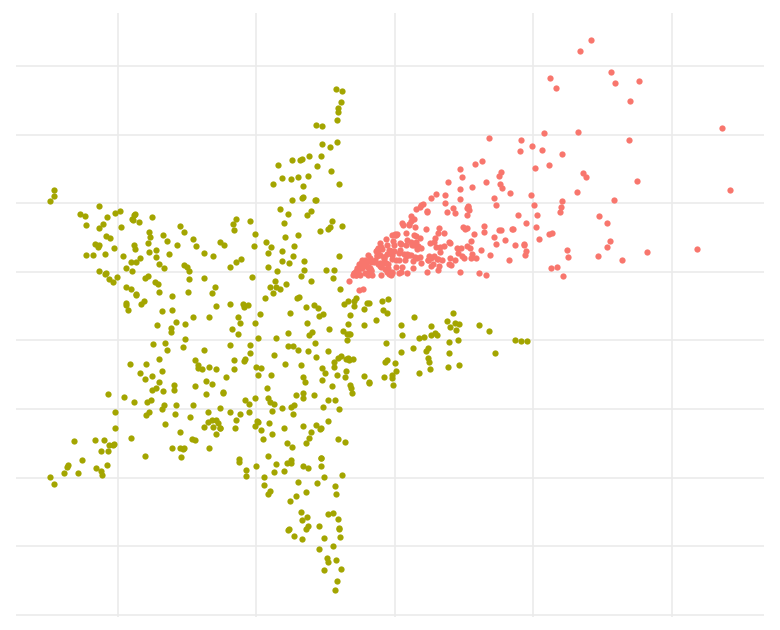}
        \caption{A-DLCC (Starbeam)}
    \end{subfigure}
    \hfill
    \begin{subfigure}[b]{0.23\textwidth}
        \includegraphics[width=\linewidth]{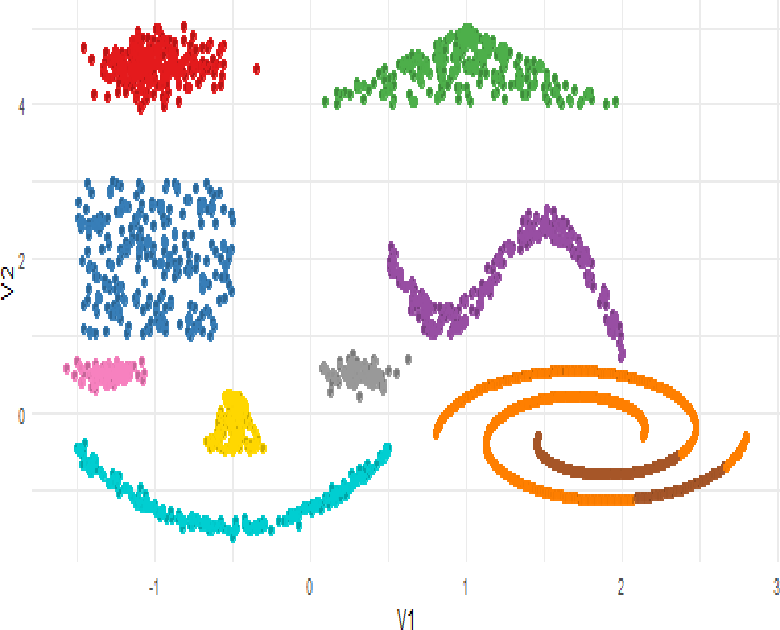}
        \caption{DLCC (Blend)}
    \end{subfigure}
    \hfill
    \begin{subfigure}[b]{0.23\textwidth}
        \includegraphics[width=\linewidth]{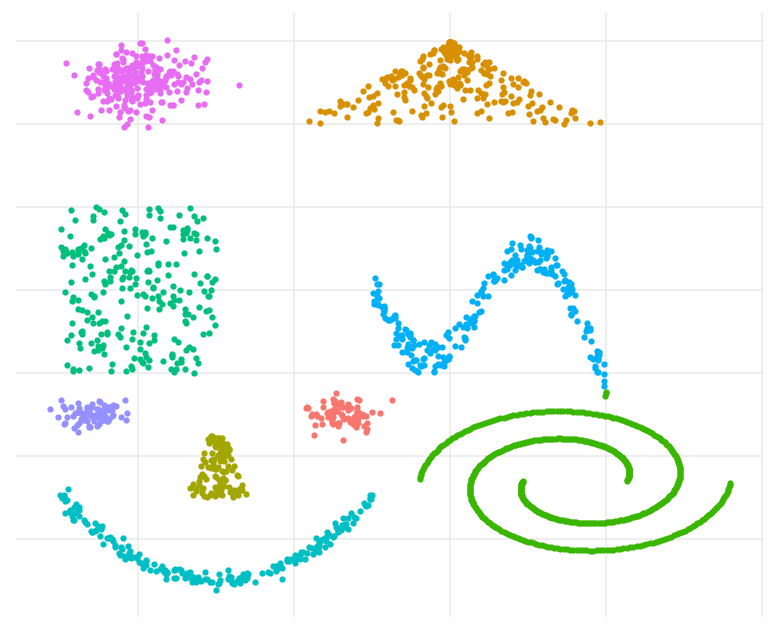}
        \caption{A-DLCC (Blend)}
    \end{subfigure}
    
    \vskip\baselineskip
    \begin{subfigure}[b]{0.23\textwidth}
        \includegraphics[width=\linewidth]{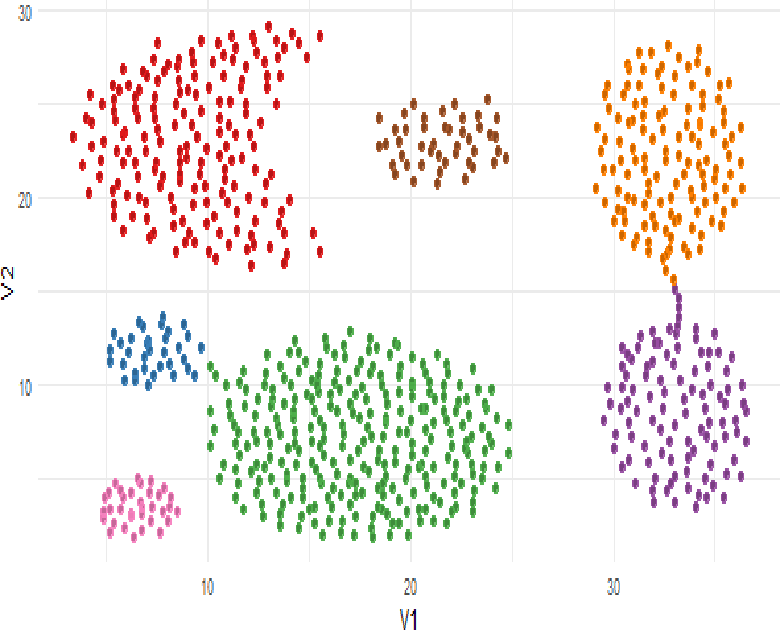}
        \caption{DLCC (Agg)}
    \end{subfigure}
    \hfill
    \begin{subfigure}[b]{0.23\textwidth}
        \includegraphics[width=\linewidth]{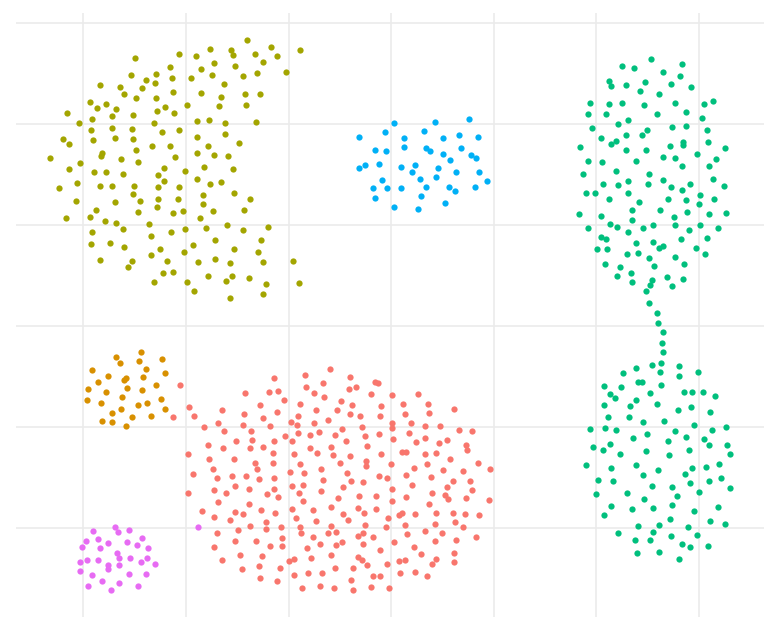}
        \caption{A-DLCC (Agg)}
    \end{subfigure}
    \hfill
    \begin{subfigure}[b]{0.23\textwidth}
        \includegraphics[width=\linewidth]{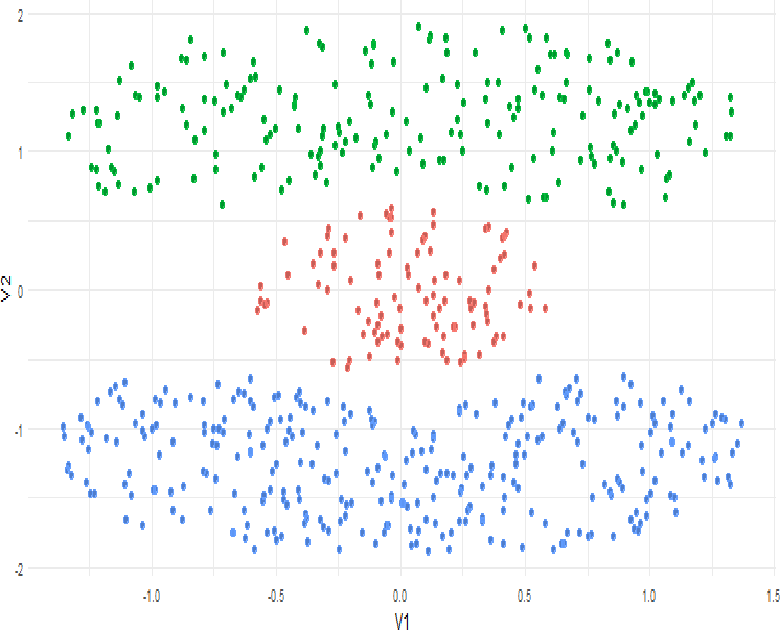}
        \caption{DLCC (Bainba)}
    \end{subfigure}
    \hfill
    \begin{subfigure}[b]{0.23\textwidth}
        \includegraphics[width=\linewidth]{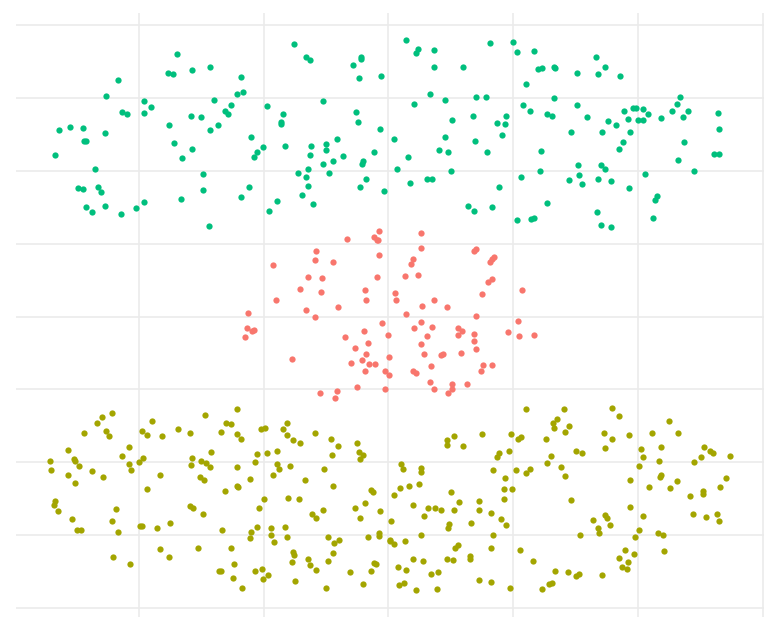}
        \caption{A-DLCC (Bainba)}
    \end{subfigure} 
    \caption{Comparison of clustering results by DLCC and A-DLCC on different synthetic datasets.}
    \label{fig:dlcc_adlcc}
\end{figure}

We evaluate A-DLCC and DLCC on four synthetic datasets: Starbeam (from~\cite{wang2025depth}), Blend and Bainba (generated using the \texttt{mlbench} R package~\cite{mlbench}), and Agg (from~\cite{gionis2007clustering}). As shown in Figure~\ref{fig:dlcc_adlcc}, A-DLCC achieves similar performance to DLCC, but without parameter tuning. There is one notable exception. The Agg dataset contains two pairs of shapes connected by a ``bridge''. DLCC separates both pairs, whereas A-DLCC separates only the pair joined by the thinner bridge (lower left) and treats the pair joined by the wider bridge (right) as a single cluster. This outcome is primarily because the merging criterion in A-DLCC is based on reachable similarity, and when there is a path connecting points in the two shapes with a density comparable to that inside the shapes, the reachable similarities between points in the two shapes are comparable to those within each shape; the disruption $\omega$ is then small on both sides and the bond~\eqref{eq:bond} is accepted. Additionally, without a fixed locality parameter, a local center can be defined within the bridge if, at a certain locality level, its neighborhood contains a balanced mix of points from both shapes. For the spiral structures in Blend, DLCC and A-DLCC also show minor differences; however, neither method fully separates the two spirals (see the discussion in~\cite{wang2025depth} for further details).

\subsection{Real data}\label{sec:rd}
We now proceed to experiments on real datasets. Specifically, we consider the eight datasets summarized in Table~\ref{tab:datasets}. Iris, Seed, Wine, Pa, BC, and Seg are widely used UCI benchmark datasets for clustering evaluation. Yale-B and Optidigits are included in the R package \texttt{PPCI}~\cite{PPCI2019}; the former is a face dataset containing $2000$ images of $10$ individuals, which is a subset from~\cite{georghiades2002few} compressed to $30 \times 20$ pixels, and the latter is a handwritten digit dataset.
\begin{table}[htbp]
\centering
\caption{Summary of the datasets}
\label{tab:datasets}
\begin{tabular}{lcccc}
\toprule
Dataset & Instances & Attributes & Clusters \\
\midrule
Iris             &   $150$   &    $4$     &  $3$ \\
Seed             &   $210$   &    $7$     &  $3$ \\
Wine             &   $178$   &   $13$     &  $3$ \\
Parkinsons (Pa)  &   $195$   &   $22$     &  $2$ \\
Breast Cancer (BC) & $569$   &   $30$     &  $2$ \\
Segmentation (Seg) & $2086$  &   $18$     &  $7$ \\
Yale-B           &  $2000$   &  $600$     & $10$ \\
Optidigits       &  $5620$   &   $64$     & $10$ \\
\bottomrule
\end{tabular}
\end{table}
\subsubsection{External Metrics}
To assess clustering performance, we adopt three widely used external evaluation metrics: adjusted Rand index (ARI)~\cite{hubert1985comparing}, normalized mutual information (NMI)~\cite{strehl2002cluster}, and purity~\cite{zhao2001criterion}.

The ARI measures pairwise similarity between clustering results and ground truth, with an expected value of $0$ under random class assignment and a value of $1$ under perfect agreement. As it penalizes both over-segmentation and under-segmentation, ARI is considered a balanced and widely adopted metric in clustering research. NMI quantifies the shared information between the predicted and true clusterings, normalized to $[0,1]$. NMI remains robust even when the number of clusters is smaller than the ground truth, but may not penalize over-segmentation as strongly as ARI. Purity measures the fraction of points in each cluster that belong to the best-matching true class, averaged over all clusters. Its value also ranges from $0$ to $1$. It is particularly informative when the number of clusters exceeds the true number of classes, as it reflects optimal matching to ground truth labels. However, purity tends to favor over-segmentation.
\subsubsection{Comparison algorithms and experimental setup}
For method comparison, we first include DLCC to compare against A-DLCC, since A-DLCC is designed as an automatic, parameter-free alternative to DLCC. Note that DLCC itself is not an automatic clustering method, and we use it as a reference method with optimally tuned parameters (assuming the true number of clusters is known) to assess whether A-DLCC can achieve comparable or even superior performance without parameter tuning. Beyond this, we include representative algorithms from several major categories: (1) mean-shift type, (2) spectral clustering with automatic estimation of the number of clusters, (3) hierarchical clustering, and (4) $K$means-type algorithms with automatic splitting and merging. Specifically, the selected methods are NN-RMS~\cite{cariou2022novel} (mean-shift), Spectrum~\cite{john2020spectrum} (spectral), FINCH~\cite{sarfraz2019efficient} (hierarchical), and U-$K$means~\cite{sinaga2020unsupervised} ($K$means type).

All these methods are capable of estimating the number of clusters, denoted by $\hat{K}$, and are compared against the ground truth $K$. However, most of them still require additional parameters to be specified for optimal performance. For NN-RMS, which requires a neighborhood parameter $k$, we follow the recommendation in~\cite{cariou2022novel} and test $k$ values from $6$ to $50$, reporting the best result. For Spectrum, we evaluate both the eigengap and multimodality gap methods for determining the number of clusters, and present the better one. For FINCH, which provides multiple clustering results at different levels, we select the result whose number of clusters is closest to the true $K$. For U-$K$means, although the algorithm is designed to be parameter-free, its objective function includes a tuning parameter $\gamma$. The original paper suggests setting $\gamma = \exp(-\hat{K}/250)$, whereas the published demo code uses $\gamma = \exp(-\hat{K}/450)$. We test both settings and report the better outcome. For methods involving randomness, we report the best result among $10$ independent runs. To avoid any ambiguity, all ``best'' or ``better'' results in this section refer to those achieving the highest ARI, which may not coincide with the best values for NMI or purity.

For A-DLCC, it is important to note that the construction of temporary clusters does not require any parameter tuning. However, the final step of the DLCC framework involves classifying the remaining observations, which may require parameters depending on the classification method used. Following the original DLCC paper, we employ two simple yet effective classification methods: depth-based $k$NN for the Seed, BC, and Optidigits datasets, and random forests for all remaining datasets. To fairly demonstrate the effectiveness of the A-DLCC framework, we do not tune the parameters of these classifiers to optimize performance; for $k$NN, the number of neighbors is fixed at $7$, and for random forests, the number of trees is fixed at $100$. Since a random forest depends on its random seed, the classification step of A-DLCC is repeated with $100$ seeds on every data set that uses random forests, and the mean over the $100$ runs is reported. 
\subsubsection{Evaluation}
Table~\ref{tab:realdata} compares the clustering performance of the A-DLCC method with DLCC, as well as several other automatic clustering algorithms, across a range of real datasets~\cite{Dua:2019,PPCI2019}. The values in parentheses indicate the percentages of points clustered and corresponding metric scores before the classification step. As shown, the temporary clusters generally achieve strong performance, justifying the last classification step. As expected, A-DLCC achieves performance comparable to DLCC, which relies on parameter selection. Notably, in some datasets, A-DLCC slightly outperforms the parameterized DLCC and even achieves  purity of $1.000$ in the Yale-B data. A-DLCC matches or exceeds the other automatic methods in both estimated cluster number and clustering accuracy, except on Seg. 

\begin{table*}[ht!]
\centering
\caption[Comparison of clustering results for real datasets.]{Comparison of clustering results for real datasets. For each dataset, the best value for each metric is shown in bold. ``--'' in $\hat{K}$ indicates the true $K$ is provided; $^*$ indicates $K$ is selected from the method's output. For A-DLCC, values in parentheses indicate the temporary clustering coverage (in the $\hat{K}$ row) and the corresponding temporary clustering ARI, Purity, and NMI (in each metric row). For the data sets classified with random forests (Iris, Wine, Pa, Seg, Yale-B), the final A-DLCC values are means over $100$ random seeds. }
\label{tab:realdata}
\resizebox{0.95\textwidth}{!}{
\begin{tabular}{llcccccc}
\toprule
Dataset & Metric & A-DLCC & DLCC & Spectrum & NN-RMS & FINCH & U-$K$means \\
\midrule
\multirow{4}{*}{Iris} 
 & $\hat{K}$ & $3$ {\scriptsize(79.3\%)}  & -- & $3$ & $3$ & $3^*$ & $3$ \\
 & ARI    & $0.846$ {\scriptsize(0.936)} & $0.878$ & $0.851$ & $0.746$ & \textbf{0.886} & $0.746$ \\
 & Purity & $0.945$ {\scriptsize(0.975)} & $0.957$  & $0.945$ & $0.900$ & \textbf{0.960} & $0.900$ \\
 & NMI    & $0.830$  {\scriptsize(0.905)}& $0.858$ & $0.832$ & $0.798$ & \textbf{0.871} & $0.798$ \\
\midrule
\multirow{4}{*}{Seed}
 & $\hat{K}$ & $3$ {\scriptsize(77.1\%)} & -- & $3$ & $3$ & $3^*$ & $3$ \\
 & ARI    & $0.762$ {\scriptsize(0.948)} & \textbf{0.775} & $0.596$ & $0.766$ & $0.701$ & $0.732$ \\
 & Purity & $0.914$ {\scriptsize(0.981)} & \textbf{0.919} & $0.838$ & $0.914$ & $0.886$ & $0.905$ \\
 & NMI    & $0.714$  {\scriptsize(0.916)} & \textbf{0.731} & $0.631$ & $0.729$ & $0.678$ & $0.723$ \\
\midrule
\multirow{4}{*}{Wine}
 & $\hat{K}$ & $3$ {\scriptsize(77.0\%)} & -- & $3$ & $3$ & $3^*$ & $3$ \\
 & ARI    & $0.911$ {\scriptsize(0.978)} & \textbf{0.930} & $0.917$ & $0.816$ & $0.686$ & $0.895$ \\
 & Purity & $0.970$ {\scriptsize(0.993)}& \textbf{0.978} & $0.972$ & $0.938$ & $0.888$ & $0.966$ \\
 & NMI    & $0.889$ {\scriptsize(0.968)}& \textbf{0.911} & $0.883$ & $0.809$ & $0.714$ & $0.865$ \\
\midrule
\multirow{4}{*}{Pa}
 & $\hat{K}$ & $2$ {\scriptsize(70.3\%)} & -- & $4$ & $2$ & $6^*$ & $2$ \\
 & ARI    & $0.397$ {\scriptsize(0.394)} & \textbf{0.422} & $0.106$ & $0.289$ & $0.036$ & $-0.098$ \\
 & Purity & $0.839$ {\scriptsize(0.839)} & $0.856$ & $0.862$ & \textbf{0.928} & $0.754$ & $0.754$ \\
 & NMI    & $0.242$ {\scriptsize(0.245)} & \textbf{0.320} & $0.292$ & $0.289$ & $0.032$ & $0.098$ \\
\midrule
\multirow{4}{*}{BC}
 & $\hat{K}$ & $2$ {\scriptsize(78.7\%)} & -- & $2$ & $3$ & $3^*$ & $2$ \\
 & ARI    & \textbf{0.786} {\scriptsize(0.964)} & $0.748$ & $0.701$ & $0.673$ & $0.647$ & $0.659$ \\
 & Purity & \textbf{0.944} {\scriptsize(0.991)} & $0.933$ & $0.919$ & $0.931$ & $0.933$ & $0.901$ \\
 & NMI    & \textbf{0.726} {\scriptsize(0.931)} & $0.655$ & $0.579$ & $0.582$ & $0.559$ & $0.546$ \\
\midrule
\multirow{4}{*}{Seg}
 & $\hat{K}$ & $12$ {\scriptsize(84.3\%)} & -- & $7$ & $12$ & $12^*$ & $4$ \\
 & ARI    & $0.254$ {\scriptsize(0.293)} & \textbf{0.589} & $0.361$ & $0.510$ & $0.479$ & $0.368$ \\
 & Purity & $0.549$ {\scriptsize(0.597)} & \textbf{0.766} & $0.569$ & $0.716$ & $0.657$ & $0.531$ \\
 & NMI    & $0.545$ {\scriptsize(0.588)}& \textbf{0.673} & $0.562$ & $0.624$ & $0.619$ & $0.581$ \\
\midrule
\multirow{4}{*}{Yale-B}
 & $\hat{K}$ & $11$ {\scriptsize(74.8\%)} & -- & $9$ & $14$ & $12^*$ & $7$ \\
 & ARI    & $0.975$ {\scriptsize(0.974)} & \textbf{0.990} & $0.757$ & $0.847$ & $0.798$ & $0.437$ \\
 & Purity & \textbf{1.000} {\scriptsize(1.000)}& $0.995$ & $0.790$ & $0.957$ & $0.845$ & $0.570$ \\
 & NMI    & $0.986$ {\scriptsize(0.986)} & \textbf{0.990} & $0.900$ & $0.895$ & $0.895$ & $0.612$ \\
\midrule
\multirow{4}{*}{Optidigits}
 & $\hat{K}$ & $13$ {\scriptsize(74.4\%)} & -- & $9$ & $14$ & $11^*$ & $3$ \\
 & ARI    & \textbf{0.834} {\scriptsize(0.901)} & $0.814$ & $0.653$ & $0.780$ & $0.000$ & $0.231$ \\
 & Purity & \textbf{0.937} {\scriptsize(0.970)} & $0.906$ & $0.741$ & $0.907$ & $0.122$ & $0.291$ \\
 & NMI    & \textbf{0.851} {\scriptsize(0.912)} & $0.839$ & $0.776$ & $0.841$ & $0.080$ & $0.451$ \\
\bottomrule
\end{tabular}
}
\end{table*}

\begin{figure}[ht!]
    \centering
    \includegraphics[width=\linewidth]{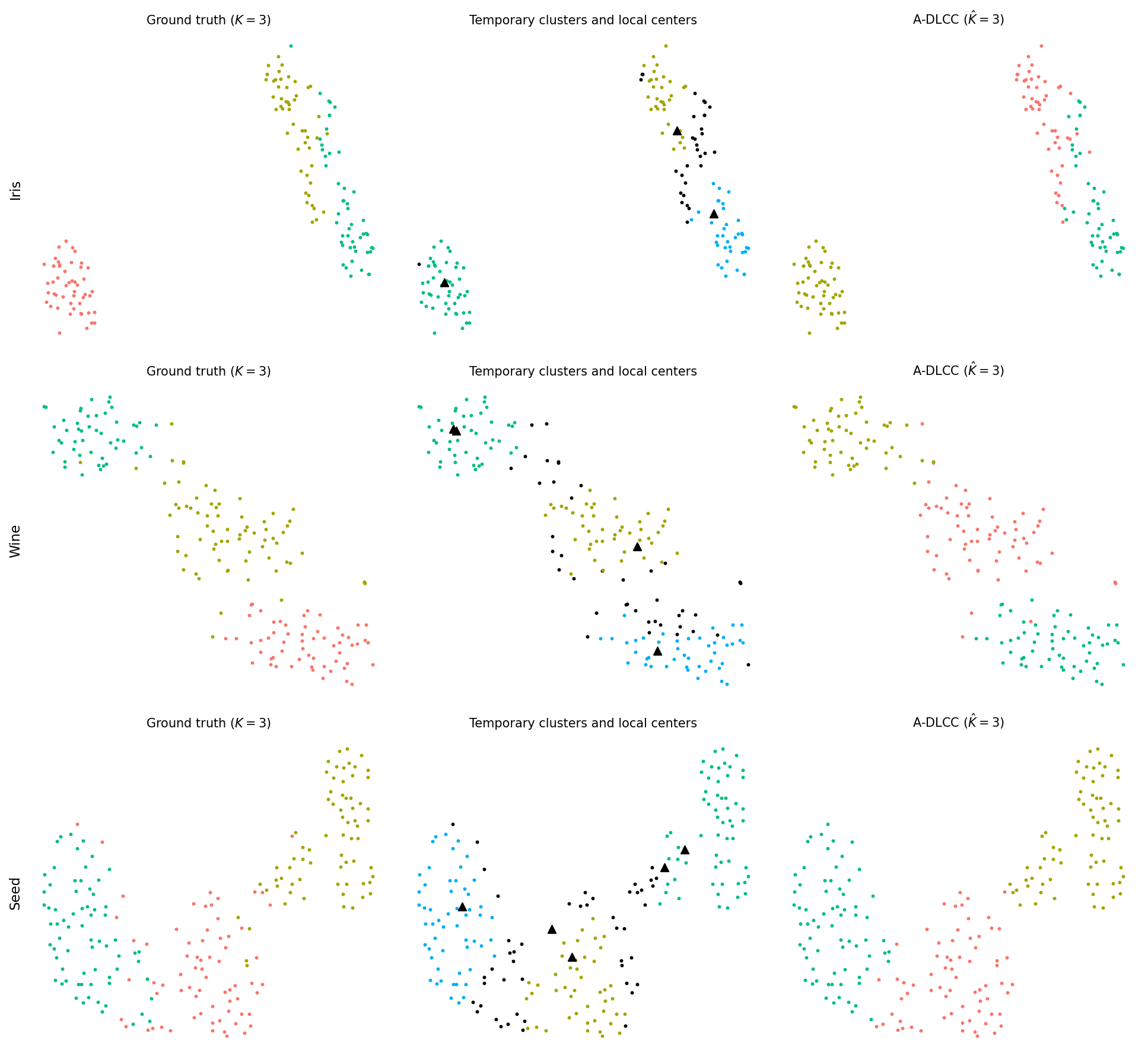}
    \caption[t-SNE visualizations for the Iris, Wine, and Seed datasets.]{t-SNE visualizations for the Iris (top), Wine (middle), and Seed (bottom) datasets. For each dataset, the three panels represent the true labels, the temporary clusters identified by A-DLCC with local centers marked by black triangles, and the final A-DLCC clustering results, respectively.}
    \label{fig:tsne_combine1}
\end{figure}

\begin{figure}[ht!]
    \centering
    \includegraphics[width=\linewidth]{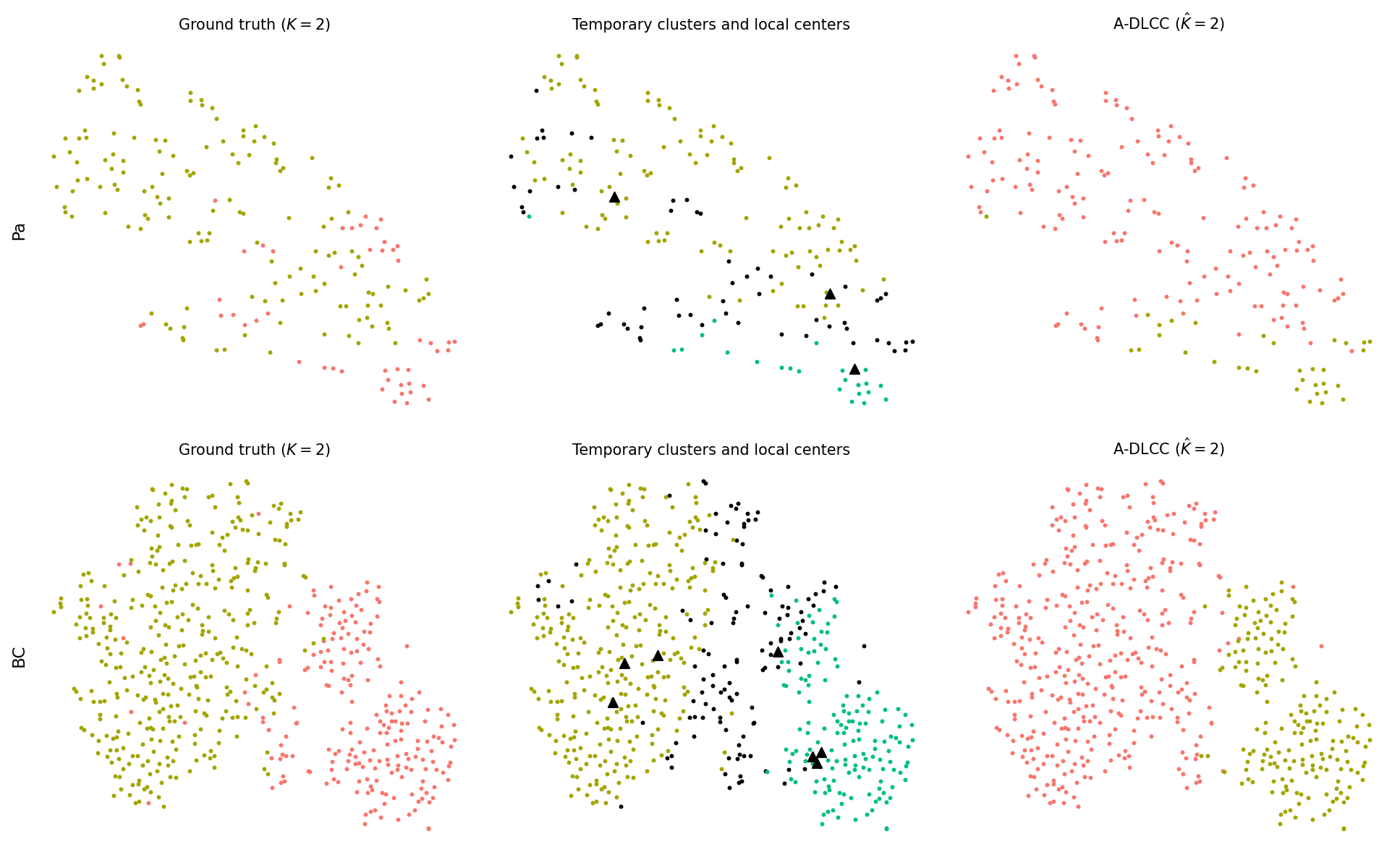}
    \caption[t-SNE visualizations for the Pa and BC datasets.]{t-SNE visualizations for the Pa (top) and BC (bottom) datasets. Panel meanings are as in Figure~\ref{fig:tsne_combine1}.}
    \label{fig:tsne_combine2}
\end{figure}

\begin{figure}[ht!]
    \centering
    \includegraphics[width=\linewidth]{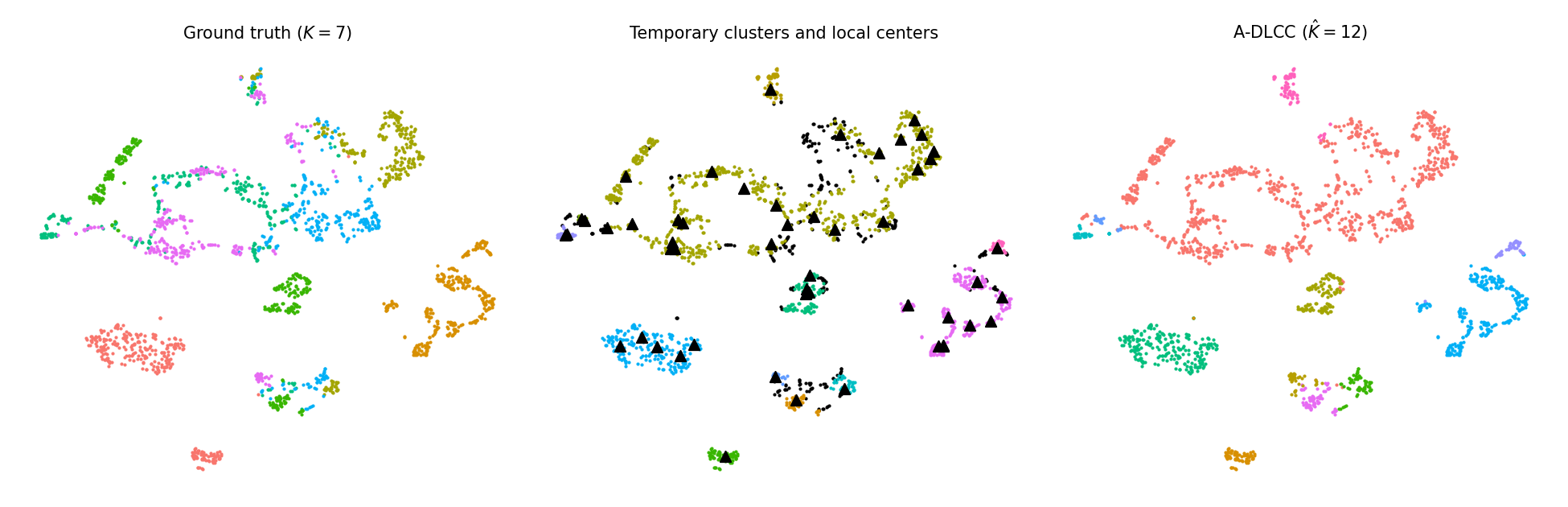}\\[2pt]
    \includegraphics[width=\linewidth]{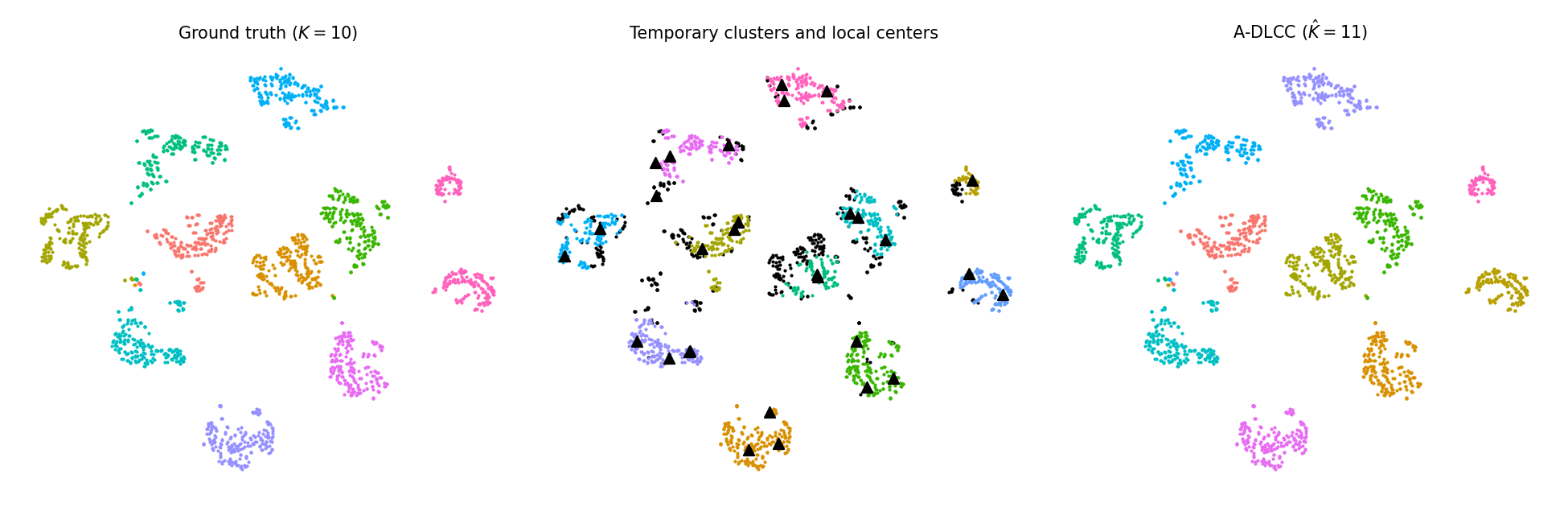}\\[2pt]
    \includegraphics[width=\linewidth]{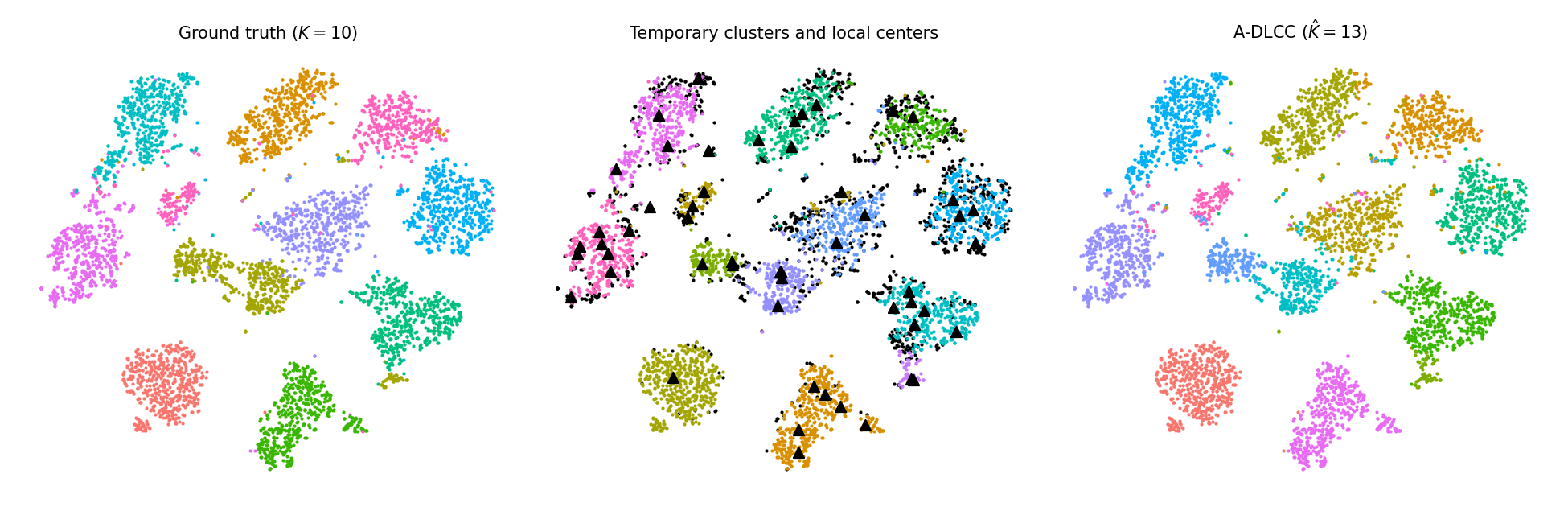}
    \caption[t-SNE visualizations for the Seg, Yale-B, and Optidigits datasets.]{t-SNE visualizations for the Seg (top), Yale-B (middle), and Optidigits (bottom) datasets. Panel meanings are as in Figure~\ref{fig:tsne_combine1}.}
    \label{fig:tsne_combine3}
\end{figure}

Figures~\ref{fig:tsne_combine1}, \ref{fig:tsne_combine2} and \ref{fig:tsne_combine3} visualize the A-DLCC clustering results alongside the ground truth labels using t-distributed stochastic neighbor (tSNE) embeddings \cite{van2008visualizing}, a nonlinear dimensionality reduction technique that projects high-dimensional data into a low-dimensional space for visualization. Figure~\ref{fig:tsne_combine1} shows datasets with clusters that are nearly visually separated and balanced in size, where the local centers identified by A-DLCC generally coincide with the geometric centers of the clusters. Figure~\ref{fig:tsne_combine2} presents the Pa and BC datasets, both exhibiting clear cluster size imbalance and some degree of overlap. Here, the unlabelled points in the temporary clustering panels are mostly found near overlapping regions and cluster boundaries. Figure~\ref{fig:tsne_combine3} highlights that for complex datasets such as Yale-B, Optidigits, and Seg, A-DLCC identifies local centers corresponding to visually separated point groups, even when these groups are relatively small. In contrast, the ground truth labels often group such small, well-separated point groups together with larger point collections. This behavior reflects the fact that A-DLCC does not assume cluster size balance when defining local centers. Seg is the exception, where several local centers fall where two or three classes overlap. The central groups are then joined because each reaches the next about as well as itself. DLCC uses a comparatively large neighborhood size and so does not define a local center in that overlap. Across all datasets, the A-DLCC clusters are visually coherent under tSNE.

\subsection{A case study on Anuran Calls}
Anuran Calls \cite{diaz2012compressive} is a widely used, highly unbalanced dataset in the clustering literature. It contains $7195$ audio syllables from $10$ frog species, each described by $22$ variables. The cluster sizes range from as few as $68$ to as many as $3478$ observations, making it a particularly challenging scenario for unsupervised algorithms.

Table~\ref{tab:frog} reports the performance of A-DLCC and other adaptive clustering baselines mentioned earlier. A-DLCC significantly outperforms all competing methods in terms of ARI and NMI, and attains the highest purity as well. It achieves an ARI above $0.9$ despite estimating a higher number of clusters ($\hat K=16$) than the ground truth. Although Spectrum yields the value of $\hat{K}$ that is the closest to the true number, it fails to detect small classes. For example, the smallest species ($n=68$) is not separated out. Methods like NN-RMS and FINCH achieve high purity but much lower ARI, since they over-split the largest class (the species with $3478$ observations) into multiple clusters.

\begin{figure}[ht!]
    \centering
    \includegraphics[width=1\linewidth]{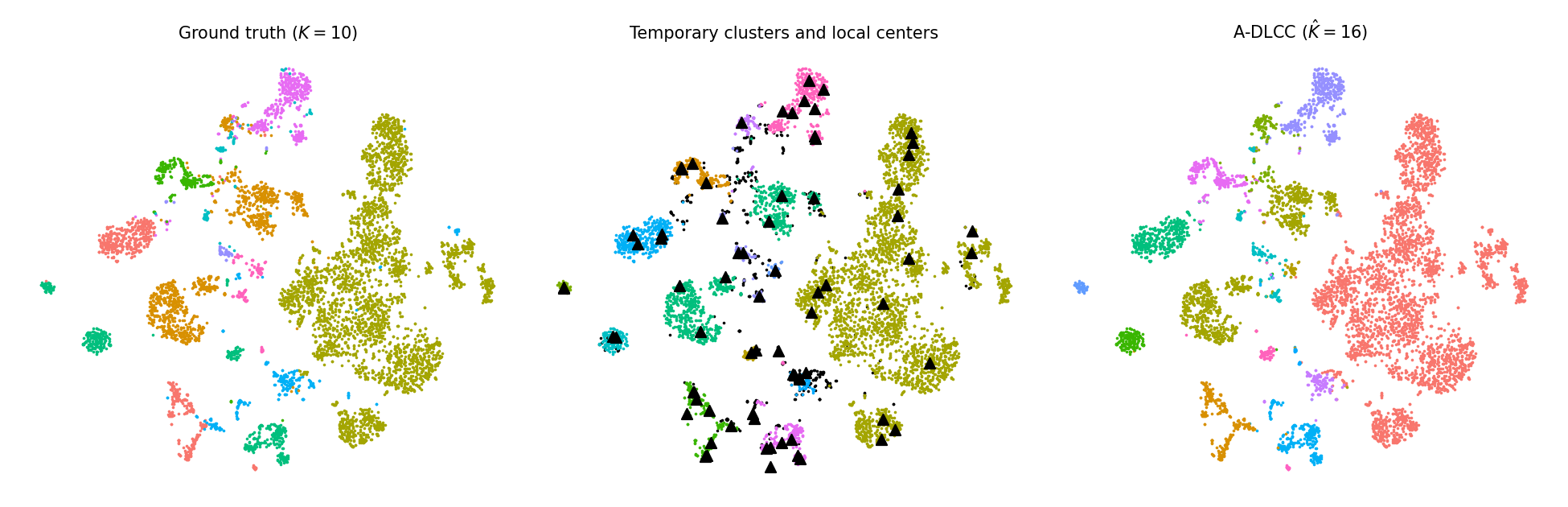}
    \caption{t-SNE visualization of Anuran Calls: ground truth, temporary clusters with local centers (triangles), and the A-DLCC result.}
    \label{fig:anuran_tsne}
\end{figure}

Figure~\ref{fig:anuran_tsne} visualizes the A-DLCC result and compares it with the ground truth. The two largest species are recovered almost perfectly: of the $3555$ observations in the largest cluster, $3468$ belong to the largest species (total $3478$ observations), and $988$ of the $1121$ observations of the second largest species fall in one cluster. The local centers of the largest species, which occupy a long, dense region of the embedding, are bonded into one group because each of them reaches its neighbors about as well as it reaches itself and the neighbors are community-level contacts on the depth graph; the same rule keeps the small species on the periphery apart, because their background level is low and their contacts with the large groups are weaker than the null model expects. The higher number of clusters comes from the medium-sized species: the species with $672$ and $542$ observations are each split into two to four clusters that correspond to visually separated point groups in the tSNE embedding, and the three smallest species ($114$, $68$ and $148$ observations) are partly gathered in one cluster. Since the large classes are pure and the splits concern comparatively few observations, the ARI remains high. A-DLCC recovers both the dominant and the smaller structures, even with a larger estimated number of clusters.

The tendency of A-DLCC to estimate a larger number of clusters in high-dimensional, complex datasets is consistent with our previous discussion. Without strong assumptions about data structure or cluster size, A-DLCC naturally identifies small groups of points that are not similar to any major clusters as separate clusters. This is a consequence of the fully automatic design.
\begin{table*}[htbp]
\centering
\caption{Comparison of clustering results on Anuran Calls; the final A-DLCC values are means over $100$ random-forest seeds.}
\label{tab:frog}
\resizebox{\textwidth}{!}{
\begin{tabular}{llccccc}
\toprule
Dataset & Metric & A-DLCC & Spectrum & NN-RMS & FINCH & U-$K$means \\
\midrule
\multirow{4}{*}{Anuran Calls} 
 & $\hat{K}$ & $16$ {\scriptsize(92.9\%)} & $7$ & $19$ & $13^*$ & $2$ \\
 & ARI    & \textbf{0.918} {\scriptsize(0.952)} & $0.590$ & $0.394$ & $0.232$ & $0.551$ \\
 & Purity & \textbf{0.927} {\scriptsize(0.963)} & $0.647$ & $0.894$ & $0.776$ & $0.636$ \\
 & NMI    & \textbf{0.806} {\scriptsize(0.872)} & $0.550$ & $0.647$ & $0.552$ & $0.597$ \\
\bottomrule
\end{tabular}
}
\end{table*}

\section{Conclusions and future directions} \label{sec:cfd}
In this paper, we proposed the A-DLCC algorithm, which does not require tuning parameters and can automatically adapt to a wide range of clustering scenarios. A-DLCC outperforms the other automatic clustering methods tested on real high-dimensional data, including cases with balanced and unbalanced clusters. Nevertheless, for some complex high-dimensional datasets, A-DLCC tends to estimate a larger number of clusters than the ground truth, particularly when the underlying structure is ambiguous or contains small, weakly connected groups. The same challenge applies to nearly all unsupervised clustering methods. Except for simple or synthetic datasets with clear cluster structure and labels, it is almost impossible to recover the ``true'' number of clusters without prior knowledge in many real-world cases. As noted by Jeon et al.~\cite{jeon2025measuring}, ground truth labels themselves may correspond to ambiguous or even arbitrary groupings, rather than a well-defined clustering. Such labels may not be reflected by the structures that are visible in the data.

For adaptive clustering methods, what matters most is that the results are meaningful and interpretable without prior knowledge of the number of clusters. Under A-DLCC, for example, ambiguous small clusters may indicate outliers within larger groups or reveal overlapping regions involving several distinct classes, and may point to targets for manual inspection or domain-specific analysis. Several limitations and open problems remain.

Computational scalability is a concern. While the original DLCC had $O(n^3)$ complexity, A-DLCC can be even more demanding. For example, as the nonparametric design considers all neighborhood sizes, computing $\beta$-ILD requires $O(n^2)$ operations, and the worst-case complexity of constructing $\mathbb{G}$ is $O(n^3)$.  Developing faster estimation methods for local depth remains an important challenge for all methodologies that rely on it.

Beyond computational issues, theoretical understanding of the merging process also deserves further investigation. The grouping rule of A-DLCC is built from quantities with a clear meaning (the relative reachability ratio, the background level, the disruption and the modularity gain), and it applies the same rule to every pair of groups without a strategy choice; nevertheless, the way the threshold~\eqref{eq:adapth} combines these quantities, and the value of the slack $q$, remain heuristic rather than being derived from a model. Ideally, one would prefer a more principled approach where merging decisions are guided by a well-defined objective function. However, to date, no universally accepted objective function exists that performs well across both convex and non-convex cluster shapes and in varying dimensionalities. Developing such an objective function or internal criterion remains an open challenge, and could support more principled model selection across different clustering algorithms.
\bibliographystyle{ieeetr}
\bibliography{references}

\end{document}